\documentclass[a4paper,11pt]{article}
\pdfoutput=1 

\usepackage{jheppub} 
\usepackage[numbers,sort&compress]{natbib} 
\usepackage[T1]{fontenc} 
\usepackage{color}

\usepackage{lineno}

\newcommand{\jpsi}{J/\psi}
\newcommand{\psip}{\psi(2S)}
\newcommand{\elp}{e^{+}}
\newcommand{\elm}{e^{-}}
\newcommand{\mup}{\mu^{+}}
\newcommand{\mum}{\mu^{-}}
\newcommand{\pip}{\pi^{+}}
\newcommand{\pim}{\pi^{-}}
\newcommand{\piz}{\pi^{0}}

\newcommand{\prp}{p}
\newcommand{\prm}{\bar{p}}

\newcommand{\kp}{K^{+}}
\newcommand{\km}{K^{-}}

\newcommand{\BESIIIorcid}[1]{\href{https://orcid.org/#1}{\hspace*{0.1em}\raisebox{-0.45ex}{\includegraphics[width=1em]{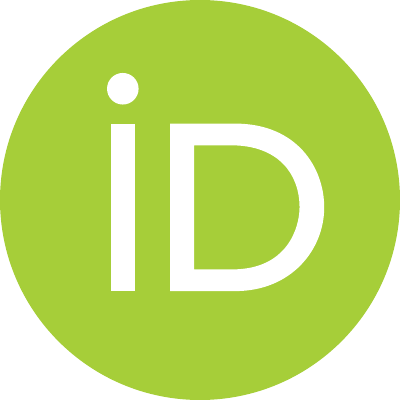}}}} 

\title{\boldmath Measurement of $\elp\elm\to\prp\prm$ around $\psip$ with scan method}

\author{The BESIII Collaboration   \\
\includegraphics[width=0.2\textwidth]{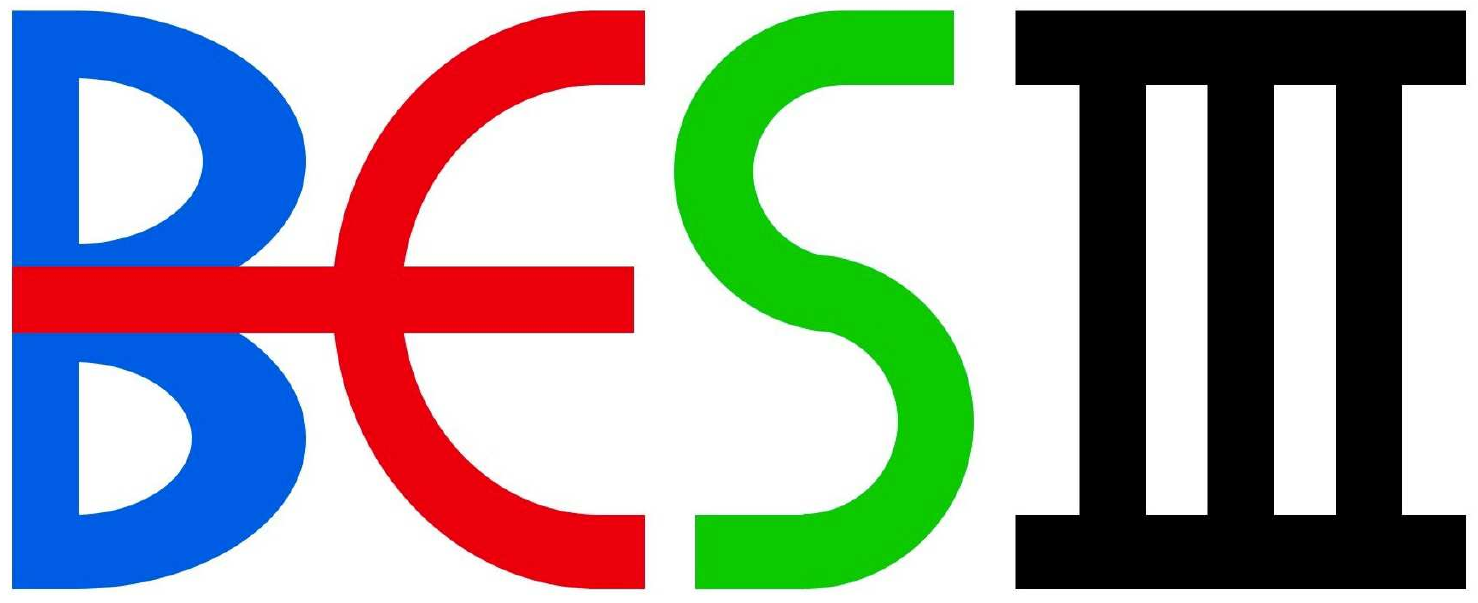}}

\begin{document}

\abstract{
This paper uses the data corresponding to an integrated luminosity of $495$~pb$^{-1}$ in the center-of-mass energy region from 3.58~GeV to 3.71~GeV to study the relative phase between strong and electromagnetic amplitudes in the $\psip\to\prp\prm$ decay.
The cross sections and the effective form factors around $\psip$ of proton-antiproton pairs are presented thereby filling the gap in the experimental results around the $\psip$ resonance for the first time, and the results are consistent with the previous BESIII and BaBar results. 
The relative phase is extracted from the cross-section line shape as $\Phi_{g,\gamma}=(105.9\pm9.4)^\circ$ or $\Phi_{g,\gamma}=(-105.3\pm9.1)^\circ$.
Correspondingly, two solutions for the $\psip\to\prp\prm$ branching fractions are obtained as $\mathcal{B}=(3.21\pm0.13)\times10^{-4}$ and $\mathcal{B}=(3.47\pm0.19)\times10^{-4}$.
These results support the universality of the phase and provide new insights into its underlying mechanism.
}

\maketitle

\flushbottom
\section{Introduction}

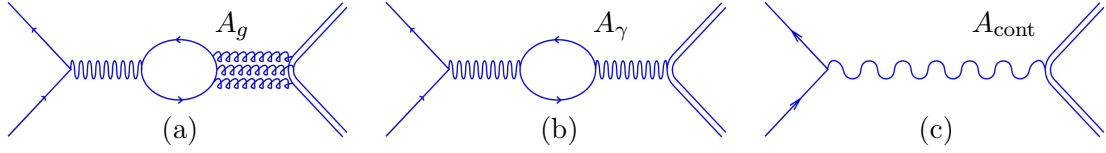
\begin{figure}[htbp]
\begin{center}
    \includegraphics[angle=0,width=4.5cm, height=1.8cm]{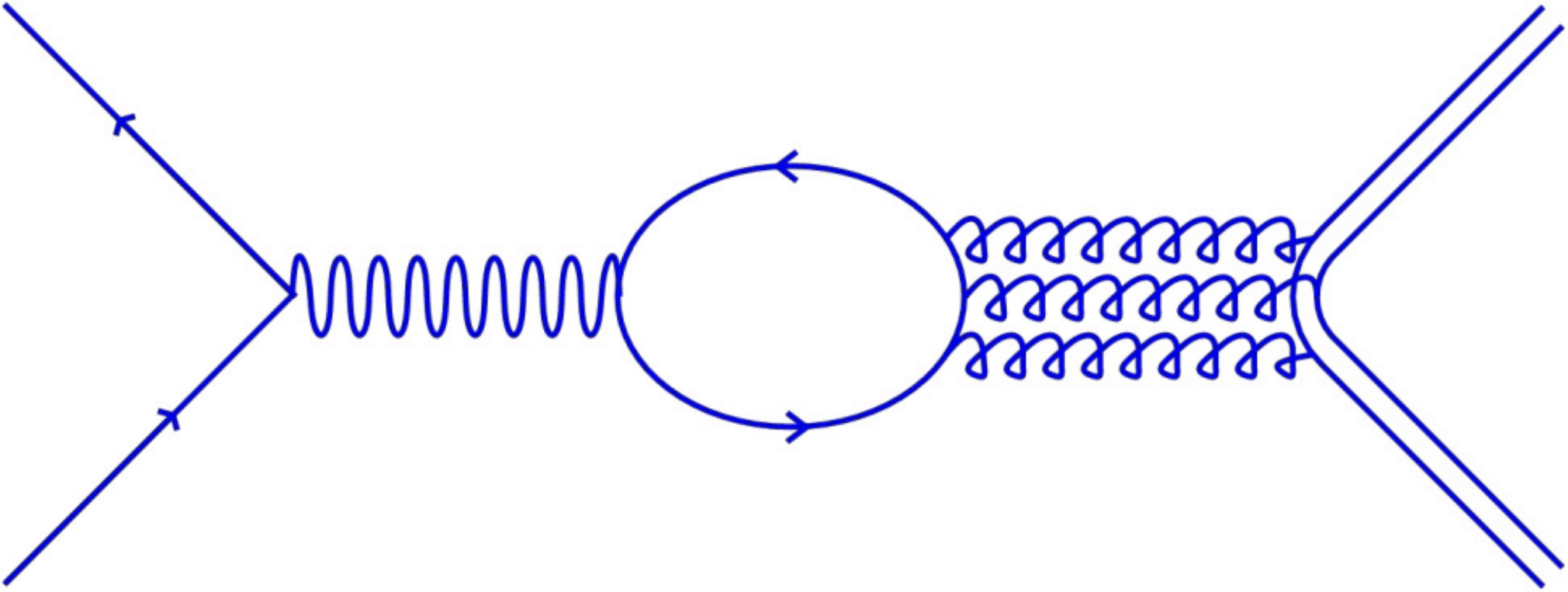}
    \put(-70, 0){(a)} \put(-50,40){$A_{g}$}
    \hskip 0.5cm
    \includegraphics[angle=0,width=4.5cm, height=1.8cm]{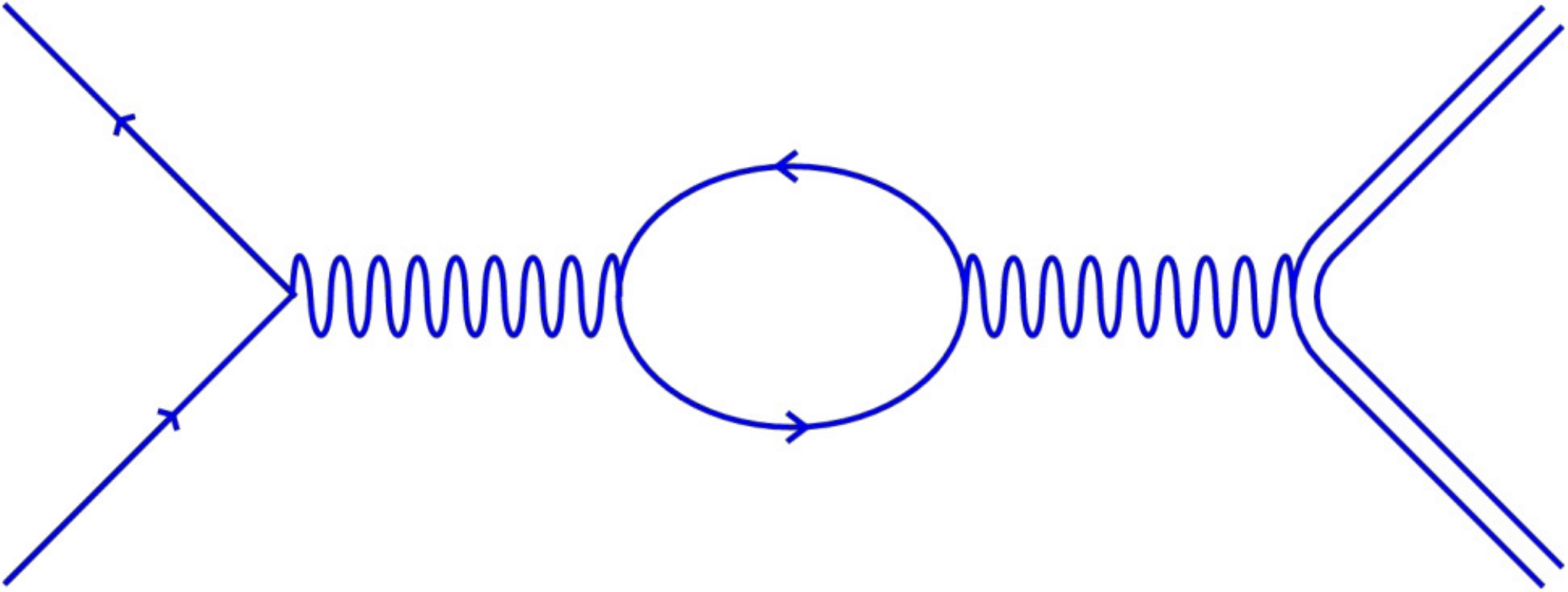}
    \put(-70, 0){(b)} \put(-50,40){$A_{\gamma}$}
    \hskip 5mm
    \includegraphics[angle=0,width=4.5cm, height=1.8cm]{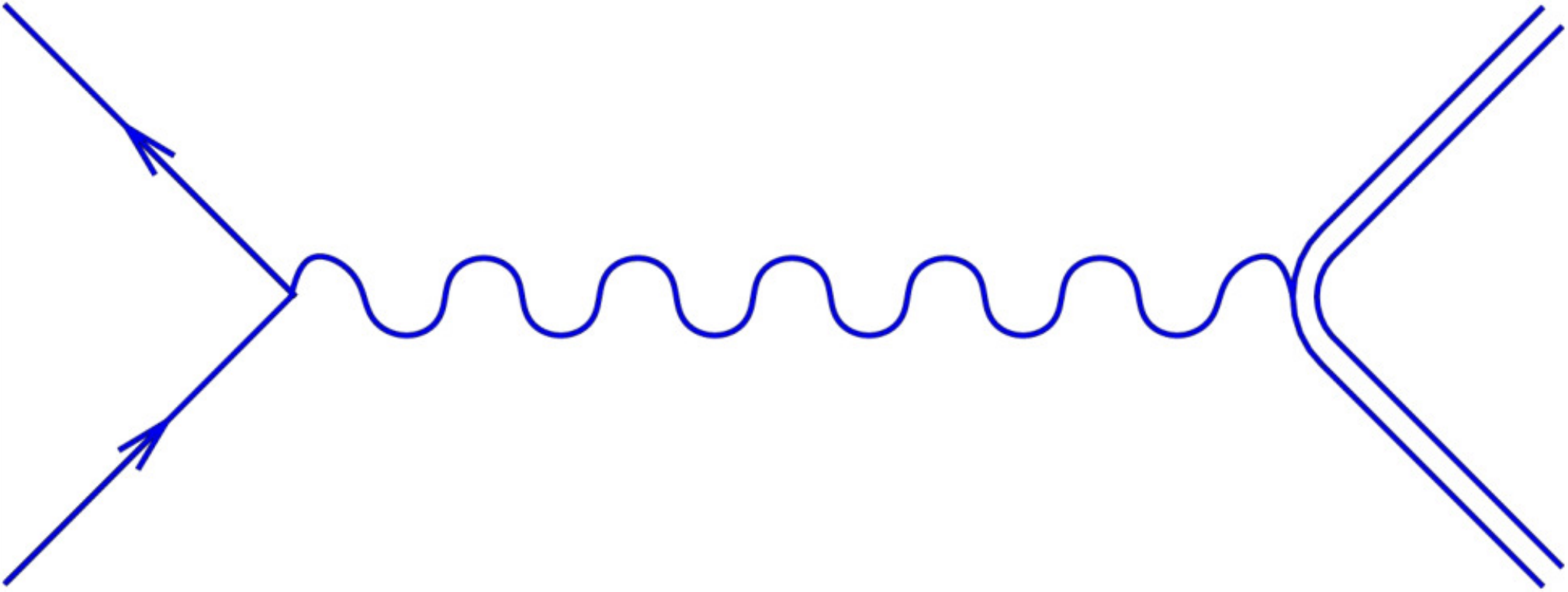}
    \put(-70, 0){(c)} \put(-50,40){$A_{\rm cont}$}
 \caption{The Feynman diagrams for the process $\elp\elm\to$ hadrons: (a) $\psip$ strong decay {\it via} gluons, (b) $\psip$ electromagnetic decay {\it via} one virtual photon, (c) the continuum process {\it via} a virtual photon.}
 \label{fig_feynman}
\end{center}
\end{figure}

 Charmonium occupies an important position in the development of Quantum Chromodynamics (QCD) due to its unique mass scale. It is approximately three times heavier than the $\phi(s\bar{s})$ meson, where non-perturbative QCD (non-pQCD) dominates the OZI-suppressed decays, yet three times lighter than the $\Upsilon(b\bar{b})$ states, where perturbative QCD (pQCD) plays the primary role~\cite{BESIII:2020nme}. The two lowest-lying $1^{--}$ states, $J/\psi$ and $\psi(2S)$, undergo OZI-suppressed hadronic decays mediated by strong interactions via three gluons ($A_g$ in figure~\ref{fig_feynman} (a)) or electromagnetic interactions via a virtual photon ($A_\gamma$ in figure~\ref{fig_feynman} (b)).

The two-body decays of $\jpsi$ are extensively studied within the SU(3) flavor symmetry framework.
For the $\jpsi$ decays to meson pairs, i.e., $1^-0^-$, $0^-0^-$, $1^-1^-$ decay modes, the branching fraction measurements from the CLEO, BES, and BESIII experiments reveal a nearly orthogonal phase between electromagnetic and strong amplitudes~\cite{LopezCastro:1994xw,BES:2003laj,Metreveli:2012tb,Suzuki:1999nb,BESIII:2025uin,Kopke:1988cs,Achasov:1999qj,Suzuki:1998ea}. As for the $\jpsi$ decays to baryon-antibaryon ($B\bar{B}$) pairs, ref.~\cite{Zhu:2015bha} obtained a similar conclusion, that the phase angle is $(-85.7\pm1.9)^\circ$ or $(90.8\pm1.6)^\circ$ taking into account the SU(3) breaking contribution. 

Within the framework of pQCD, where the coupling constants in both electromagnetic and strong interaction amplitudes are strictly real, the relative phase between these amplitudes is theoretically constrained to either $0^\circ$ or $180^\circ$. Moreover, theoretical studies~\cite{Gerard:1999uf,Wang:2004kf,Wang:2006zzg} suggest that the observed relative phase may instead manifest a universal incoherence phenomenon, particularly in charmonium decays where the three-gluon process is governed by the charm quark mass scale. Intriguingly, for $\psip$ decays, the phase proves to be compatible with $0^\circ$ within experimental uncertainties for both $1^-0^-$ and $1^+0^-$ decay modes~\cite{Suzuki:2000yq}, providing key insights into the $\rho\pi$ and non-$D\bar{D}$ puzzles
~\cite{Suzuki:2000yq,Mo:2006cy}. 
This stands in striking contrast to the $\psip\to B\bar{B}$ decay channel, where the reported phase angles of $(-98\pm25)^\circ$ or $(134\pm25)^\circ$~\cite{Zhu:2015bha} reveal significant discrepancies from $0^\circ$. These conflicting results highlight a critical need for precise phase measurements in $\psip$ decays to resolve the current theoretical ambiguities and further constrain the underlying dynamics.

Experimentally, a non-negligible continuum amplitude $A_{\rm cont}$ arises from $\elp\elm$ annihilation via a virtual photon, as depicted in figure~\ref{fig_feynman} (c). The interference between this continuum amplitude and the resonance amplitudes induces significant deviations in the measured branching ratios from their true values~\cite{Wang:2002np,BaBar:2015lgl,Guo:2022gkg}. Consequently, the interference pattern between strong and electromagnetic (EM) amplitudes can be extracted from the cross-section line shape near the charmonium resonances.

This paper employs a cross-section scan method around the $\psip$ resonance to measure both the relative phase $\Phi_{g,\gamma}$ and the branching fraction of the $\psip \to \prp\prm$ decay. The full cross section incorporating all three amplitudes ($A_g$, $A_\gamma$, $A_{\rm cont}$) for $\elp\elm\to\prp\prm$ around the $\psip$ resonance is given by
\begin{eqnarray}
    \sigma^{0}(s)=\left|A_{g}(s)e^{i\Phi_{g,\gamma}}+A_{\gamma}(s)+A_{\rm cont}(s)\right|^{2},
\end{eqnarray}
where $s$ denotes the square of the center-of-mass~(CM) energy. The relative phase between $A_{\gamma}$ and $A_{\rm cont}$ is assumed to be zero and this is verified in $\jpsi\to\mup\mum$ and $\jpsi\to\eta\pip\pim$ processes~\cite{BESIII:2018wid}.
Specifically, the total Born cross section can be expressed as
\begin{eqnarray}
\sigma(s)=\frac{4\pi\alpha^2\beta C}{3s}\left(\frac{G_0}{s^2}\right)^{2} \left( 1+\frac{1}{2\tau}\right)\left|1 + \left(1+ {\cal C} e^{i\Phi_{g,\gamma}}\right)\frac{s}{M_{\psi}}\frac{3\Gamma_{ee}/\alpha}{s-M_{\psi}^2+i\Gamma_{\psi}M_{\psi}}\right|^2\,,
\label{eq:xs1}
\end{eqnarray}
where $\alpha=1/137$ is the fine structure constant, $\beta=\sqrt{1-1/\tau}$ is the velocity of the proton with $\tau=s/4m^2_{p}$, $m_p$ is the proton mass, $\Gamma_{ee}$ is the partial width of $\psi(nS)\to\elp\elm$, $M_{\psi}$ and $\Gamma_{\psi}$  are the mass and width of the $\psi(nS)$ state. The Coulomb factor $C=y/(1-e^{-y})$, with $y=\pi\alpha/\beta$, accounts for the EM interaction between the outgoing particles. 
The term $|G_{\rm eff}|=G_{0}/s^{2}$ represents the effective form factor which was reported in ref.~\cite{BESIII:2019hdp,BESIII:2021rqk,BESIII:2019tgo}, and $G_0$ is a constant. 

The Electromagnetic form factors (EMFFs) are fundamental quantities that describe the internal structure of hadrons. 
The effective form factor $|G_{\rm eff}|$ characterizes the averaged contribution of the proton’s electric and magnetic effects to the cross section.
Recent measurements of $|G_{\rm eff}|$ reveal oscillatory modulations~\cite{BESIII:2019hdp,BESIII:2021rqk,BESIII:2019tgo,Huang:2021xte}, potentially arising from the substructure of the proton~\cite{Bianconi:2015owa,Tomasi-Gustafsson:2022tpu}, meson excitations~\cite{deMelo:2008rj,deMelo:2005cy,Lorenz:2015pba}, threshold effects~\cite{Lorenz:2015pba}, or final-state interactions~\cite{Yang:2022qoy,Qian:2022whn}. The interference between isoscalar ($I=0$) and isovector ($I=1$) EMFF components, which can be accessible through $\elp\elm\to B\bar{B}$ processes around charmonia, provides crucial insights into the isospin decomposition~\cite{Dai:2023vsw,Cao:2021asd}.

\section{The experiment and datasets}

The BESIII detector~\cite{BESIII:2009fln} records symmetric $e^+e^-$ collisions 
provided by the BEPCII storage ring~\cite{Yu:2016cof}
in the center-of-mass energy range from 1.84 to 4.95~GeV,
with a peak luminosity of $1.1 \times 10^{33}\;\text{cm}^{-2}\text{s}^{-1}$ 
achieved at $\sqrt{s} = 3.773\;\text{GeV}$. 
The BESIII detector covers 93\% of the full solid angle and consists of a helium-based
 multilayer drift chamber~(MDC), a time-of-flight
system~(TOF), and a CsI(Tl) electromagnetic calorimeter~(EMC),
which are all enclosed in a superconducting solenoidal magnet
providing a 1.0~T magnetic field.
The solenoid is supported by an
octagonal flux-return yoke with resistive plate counter muon
identification modules interleaved with steel. 

The charged-particle momentum resolution at $1~{\rm GeV}/c$ is
$0.5\%$, and the ${\rm d}E/{\rm d}x$ resolution is $6\%$ for electrons
from Bhabha scattering. The EMC measures photon energies with a
resolution of $2.5\%$ ($5\%$) at $1$~GeV in the barrel (end cap)
region. The time resolution in the plastic scintillator TOF barrel region is 68~ps, while that in the end cap region was 110~ps. The end cap TOF
system was upgraded in 2015 using multigap resistive plate chamber
technology, providing a time resolution of 60~ps~\cite{Li:2017jpg,Guo:2017sjt,Cao:2020ibk}.
All the datasets used in this analysis were taken after the TOF upgrade.

The analysis is performed based on data samples collected at nine CM energies ranging from 3.581 to 3.71~GeV. The integrated luminosities of these data sets, for a total of $495$~pb$^{-1}$ (listed in table~\ref{table_xs}), are measured using $\elp\elm\to\gamma\gamma$ events following the procedure in ref.~\cite{BESIII:2017lkp}. The CM energy $\sqrt{s}$ and its uncertainty for each data set are measured with the Beam Energy Measurement System (BEMS)~\cite{Abakumova:2011rp} with a systematic uncertainty of $2\times 10^{-5}$~\cite{Abakumova:2011rp}.

Monte Carlo~(MC) samples for signal and background channels are simulated using a {\sc geant4}-based~\cite{GEANT4:2002zbu} software package, the BESIII Object Oriented Simulation Tool~\cite{Belov:2016hxa}. The simulation models the beam
energy spread and initial state radiation (ISR) in the $e^+e^-$
annihilations with the {\sc conexc} event generator~\cite{Ping:2013jka}.
The signal process $\elp\elm\to\prp\prm$ is generated with {\sc conexc}, which includes next-to-leading order radiative corrections.
Potential backgrounds from the $\elp\elm\to\elp\elm$, $\mup\mum$, $\pip\pim$ processes are simulated with the BABAYAGA generator~\cite{CarloniCalame:2003yt}, and the background from $\kp\km$ is simulated with {\sc conexc}.
The inclusive hadronic decays of continuum processes $\elp\elm\to q\bar{q}$, with $q=u,d,s$, are simulated with the LUARLW generator~\cite{BESIII:2021wib}.
Possible backgrounds are also studied with an inclusive $\psi(2S)$ MC sample incorporated by KKMC~\cite{Jadach:2000ir}, where all particle decays are modeled with BESEVTGEN using branching fractions from Particle Data Group (PDG)~\cite{ParticleDataGroup:2024cfk} when available and LUNDCHARM otherwise~\cite{Chen:2000tv}.

\section{Event selection and background analysis}

For the $\psip \to \prp\prm$ decay, the candidates are required to have two charged tracks with zero net charge reconstructed in the MDC. The point of closest approach to the interaction point for each of the track is required to lie within a 1~cm radius in the plane perpendicular to the beam and $\pm10$~cm along the beam direction.
The polar angle of the track with respect to the $z$-axis (the symmetry axis of the MDC), $\theta$, must be inside in the fiducial volume of the barrel MDC, i.e., $\vert\!\cos\theta\vert < 0.8$, to reduce the contamination from Bhabha and dimuon events. 
In addition, the ratio $E/P$, with the deposited energy $E$ measured in the EMC and the momentum $P$ measured in the MDC, is required to be smaller than $0.5c$ for the proton candidate.  
Particle identification (PID) is performed by combining the specific ionization energy loss ${\rm d}E/{\rm d}x$ in the MDC and the flight time in the TOF. For each track, the probability for the proton particle hypothesis is required to be larger than that for the other particle hypotheses. 
Cosmic-ray background is rejected by requiring $\Delta T\equiv |T_{\prp}-T_{\prm}|<3$~ns, where $T_{\prp}$ and $T_{\prm}$ are the measured flight time for $\prp$ and $\prm$ by the TOF.
Finally, events with multiple tracks are rejected by requiring $\theta_{\prp\prm}\in[178^{\circ},180^{\circ}]$, where $\theta_{\prp\prm}$ is the opening angle between $\prp$ and $\prm$ in the $e^+e^-$ CM system.

\begin{figure}[htbp]
    \includegraphics[angle=0,width=5.0cm]{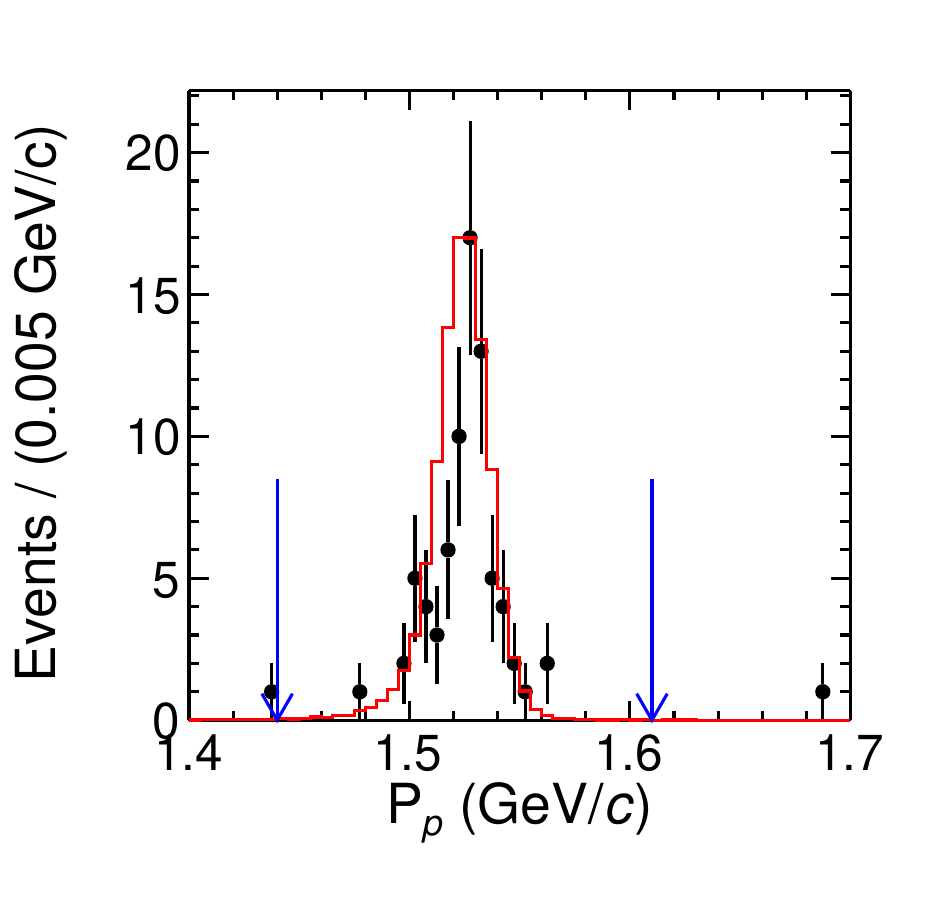} \put(-112, 90)
    {\scriptsize $3581.5$~MeV}
    \includegraphics[angle=0,width=5.0cm]{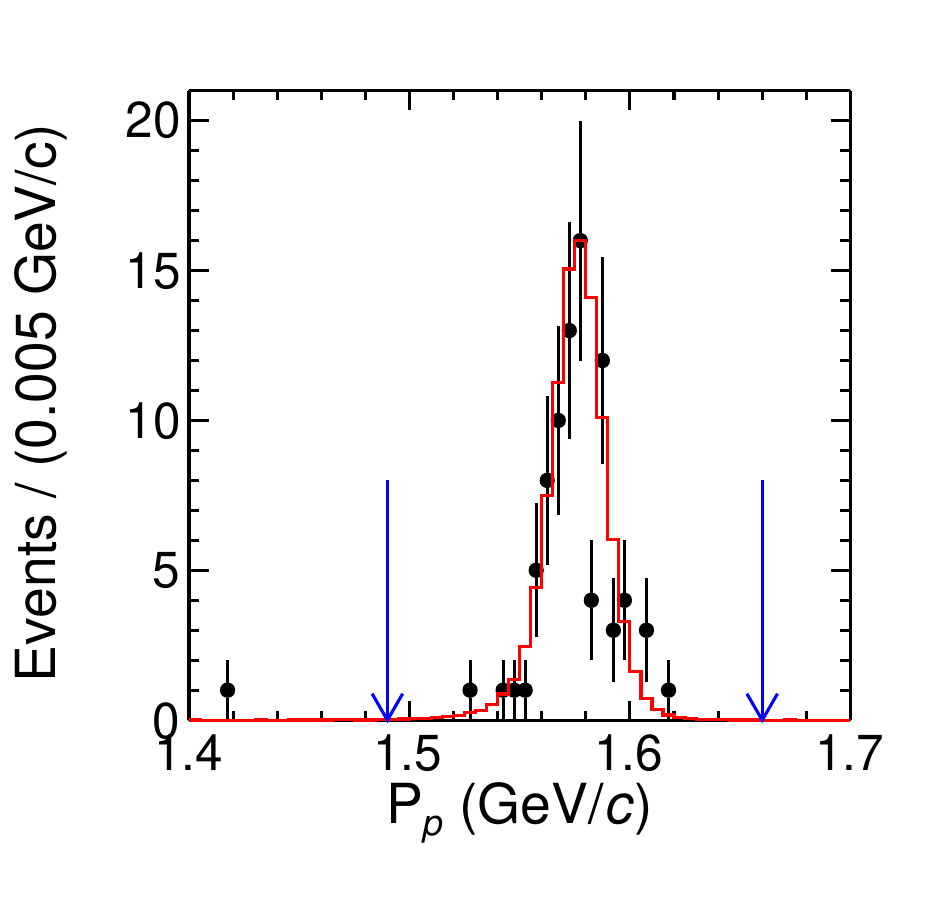} \put(-100, 90)
    {\scriptsize $3670.2$~MeV}
    \includegraphics[angle=0,width=5.0cm]{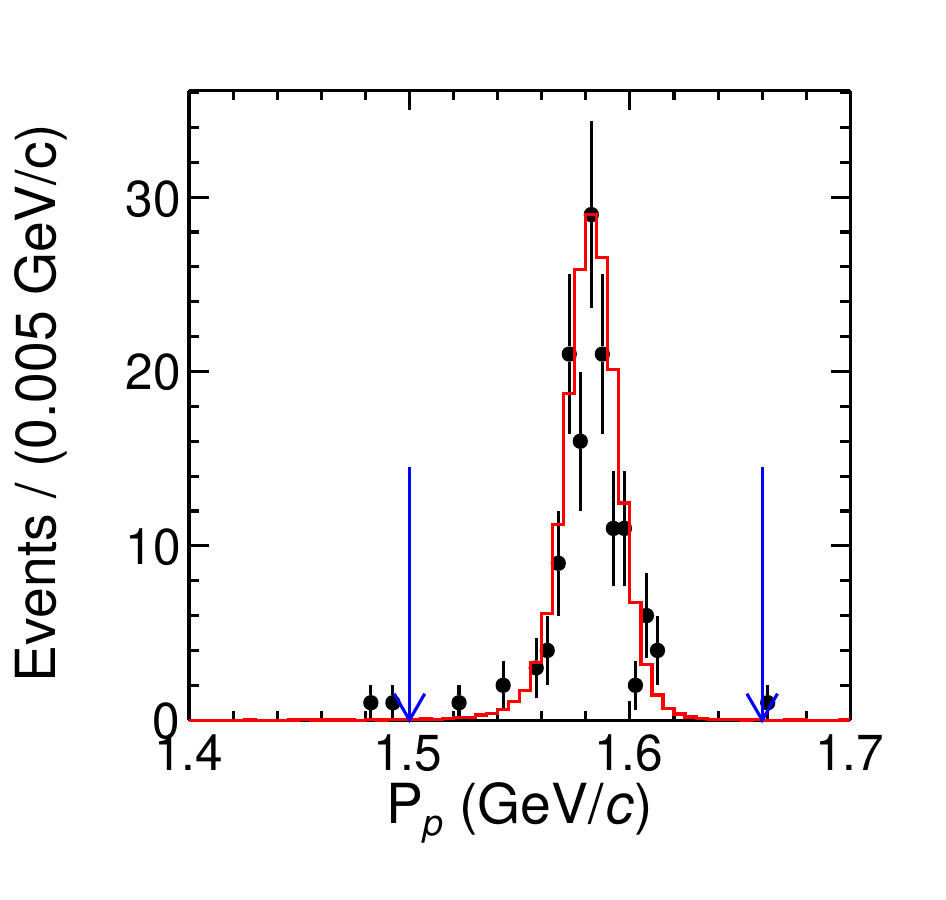} \put(-100, 90)
    {\scriptsize $3680.1$~MeV} \\
    \includegraphics[angle=0,width=5.0cm]{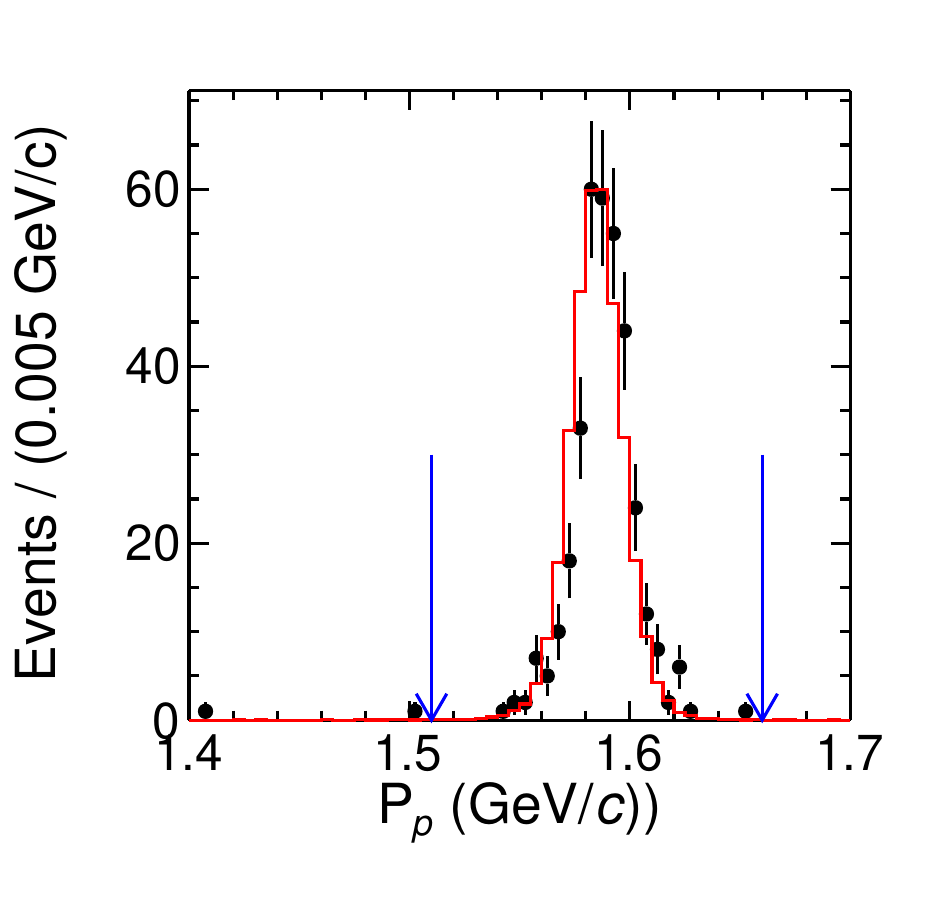} \put(-100, 90)
    {\scriptsize $3682.8$~MeV}
    \includegraphics[angle=0,width=5.0cm]{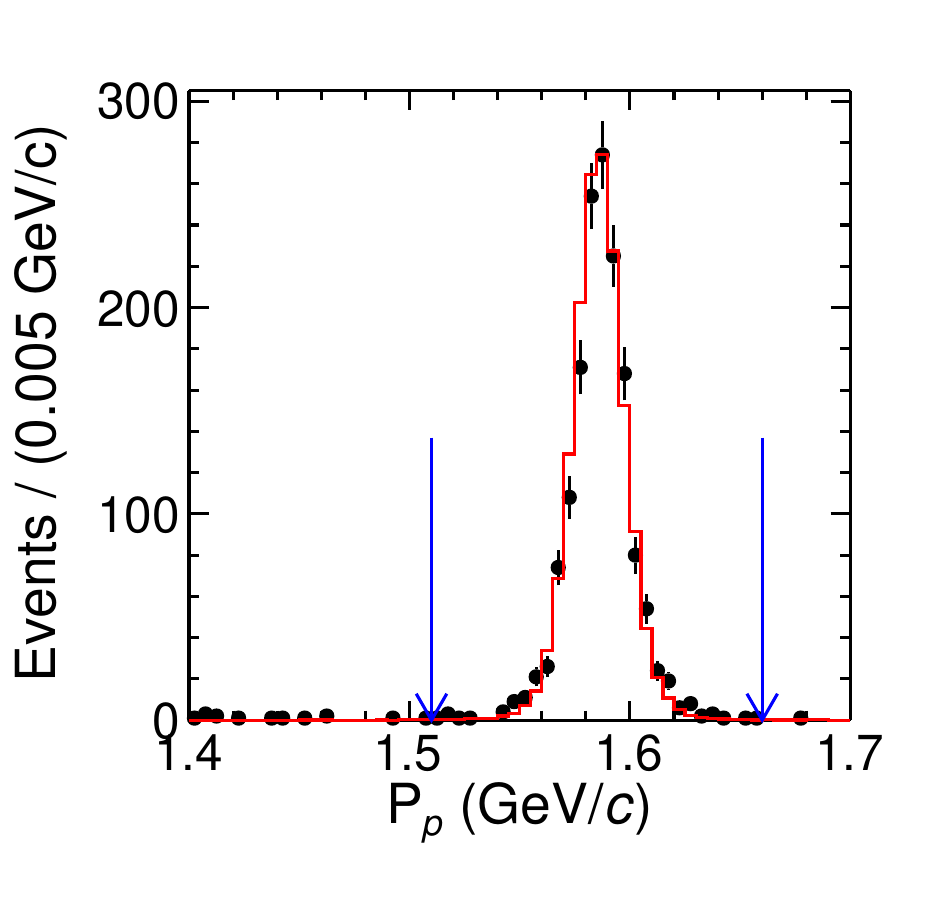} \put(-100, 90)
    {\scriptsize $3684.2$~MeV}
    \includegraphics[angle=0,width=5.0cm]{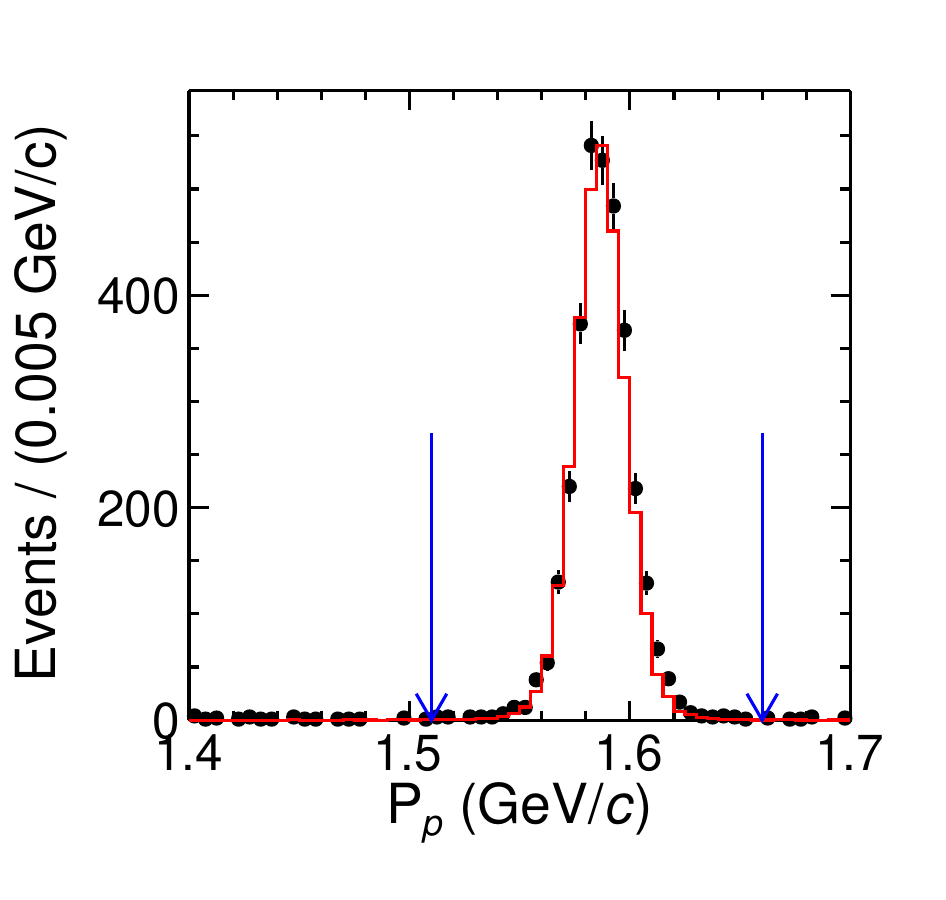} \put(-100, 90)
    {\scriptsize $3685.3$~MeV} \\
    \includegraphics[angle=0,width=5.0cm]{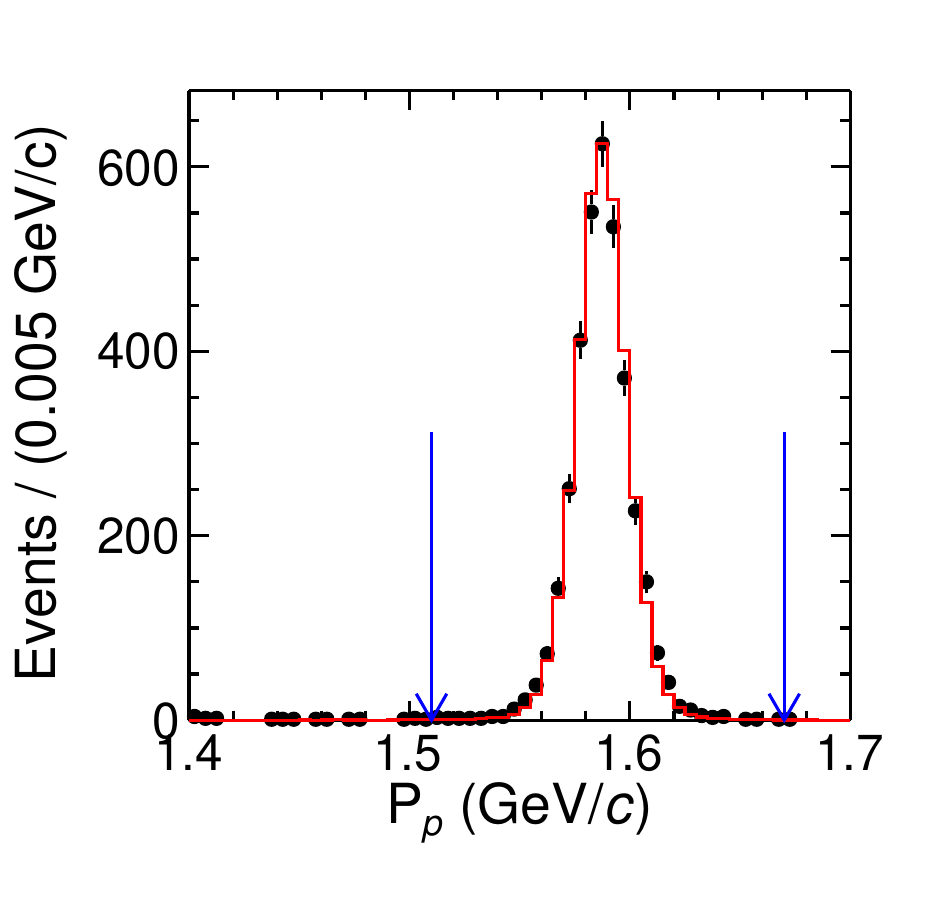} \put(-100, 90)
    {\scriptsize $3686.5$~MeV}
    \includegraphics[angle=0,width=5.0cm]{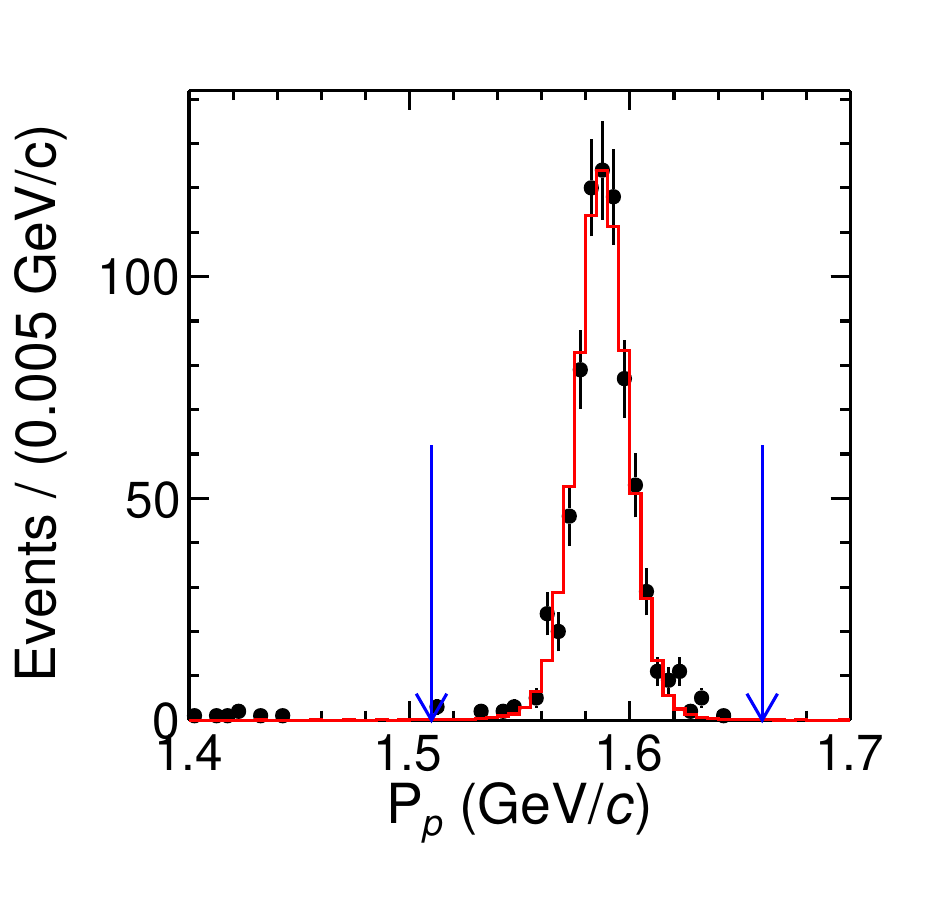} \put(-100, 90)
    {\scriptsize $3691.4$~MeV}
    \includegraphics[angle=0,width=5.0cm]{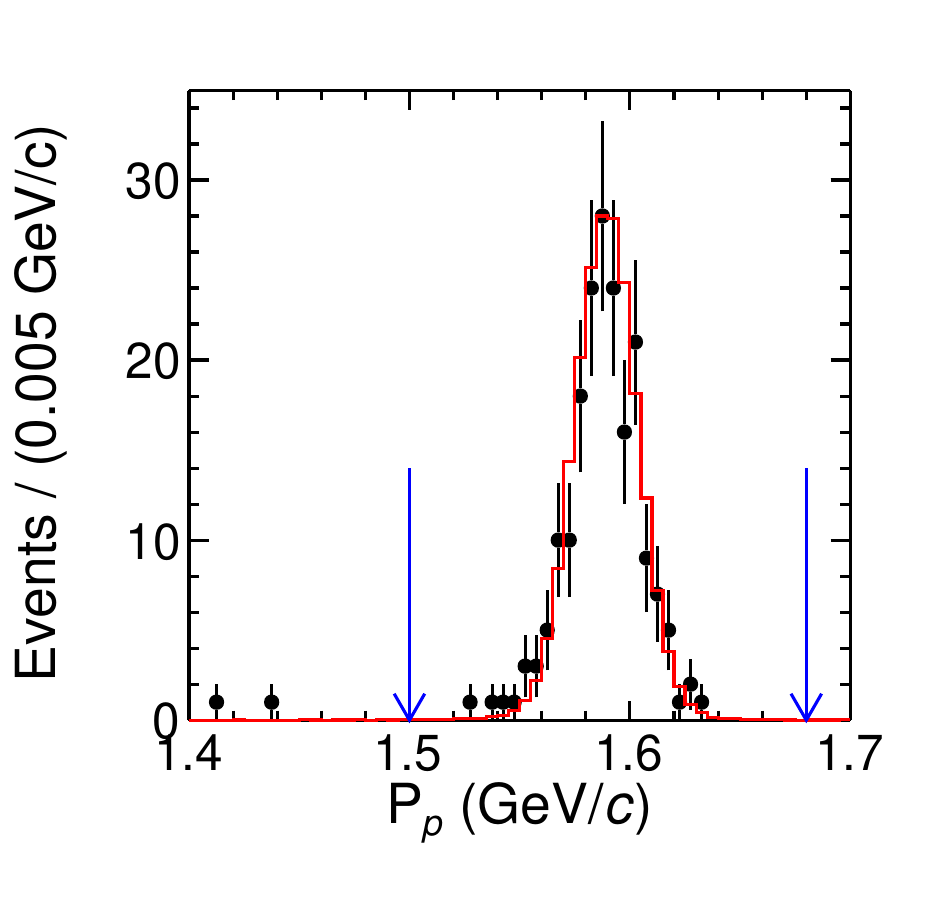} \put(-100, 90)
    {\scriptsize $3709.8$~MeV}
    \caption {Momentum distributions of proton in data and signal MC after all selection criteria. The black points are from data, the red histograms are from signal MC. The blue arrows are the symmetric momentum windows when the numbers of signal events  are extracted by counting events.}
    \label{fig_p}
\end{figure}

After applying the above selection criteria, the proton momentum $P_p$ distributions for all energy points are shown in figure~\ref{fig_p}, where the signal can be clearly distinguished, with less than 1\% background contribution which is negligible. Potential backgrounds come from hadronic processes
with multi-hadron final states $q\bar{q}$ and $\elp\elm$ annihilation into two-body final states, e.g., $\elp\elm$, $\mup\mum$, $\pip\pim$ and $\kp\km$. Dedicated background analysis is performed using MC samples ranging from 0.1 million to 10 million events for different processes, depending on the size of data samples. All backgrounds are effectively eliminated by the event selection. 

\begin{figure}[htbp]
\includegraphics[angle=0,width=5.cm]{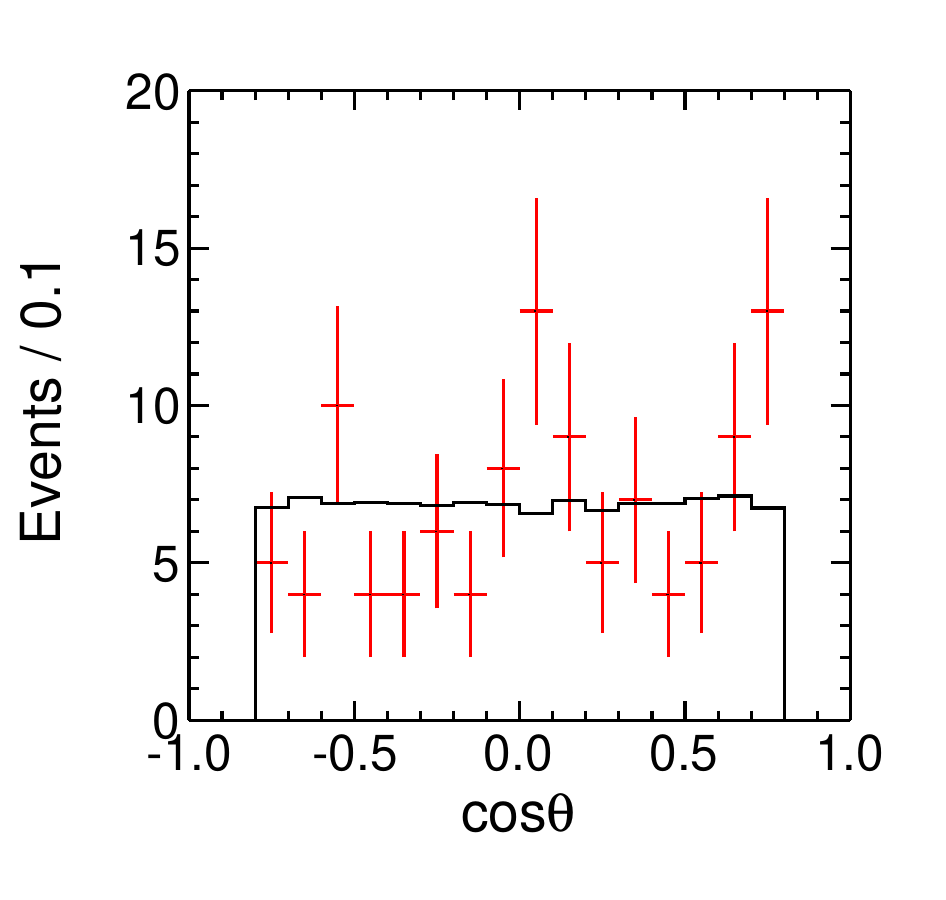}\put(-90,110){\scriptsize $3581.5$~MeV}\put(-90,100){\scriptsize $\chi^{2}/\rm {ndf}=1.32$}
\includegraphics[angle=0,width=5.0cm]{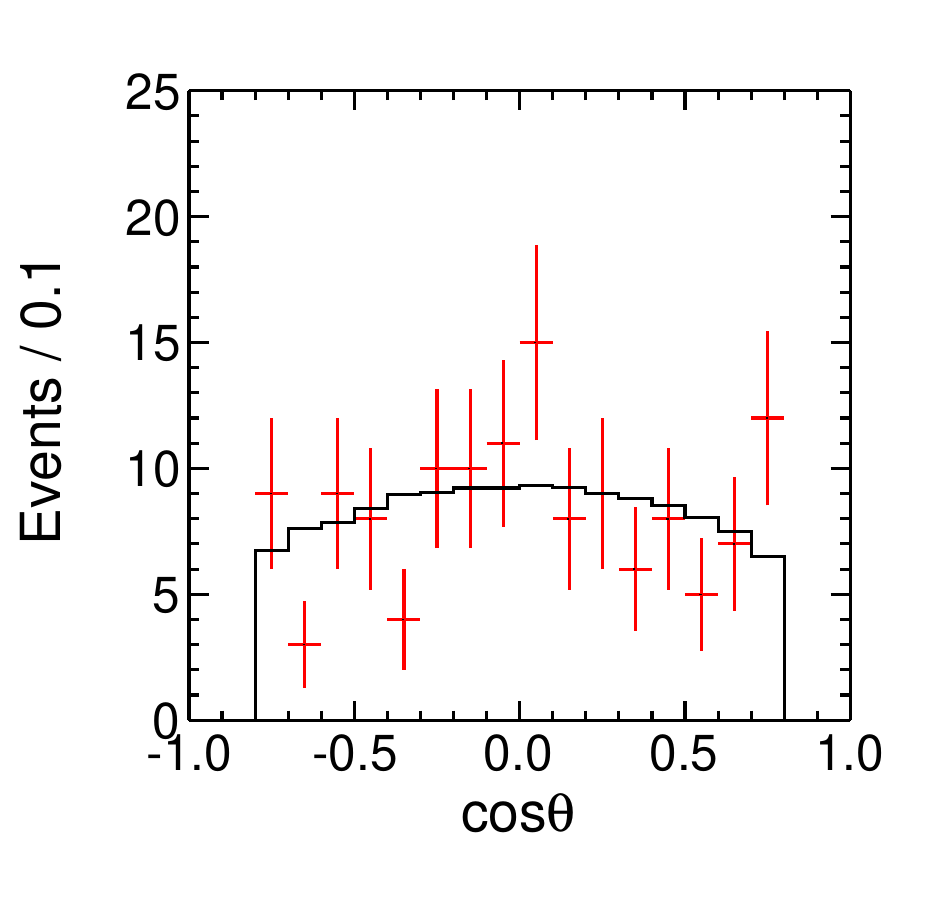}\put(-90,110){\scriptsize $3670.2$~MeV}
  \put(-90,100){\scriptsize $\chi^{2}/{\rm ndf}=1.40$}
\includegraphics[angle=0,width=5.0cm]{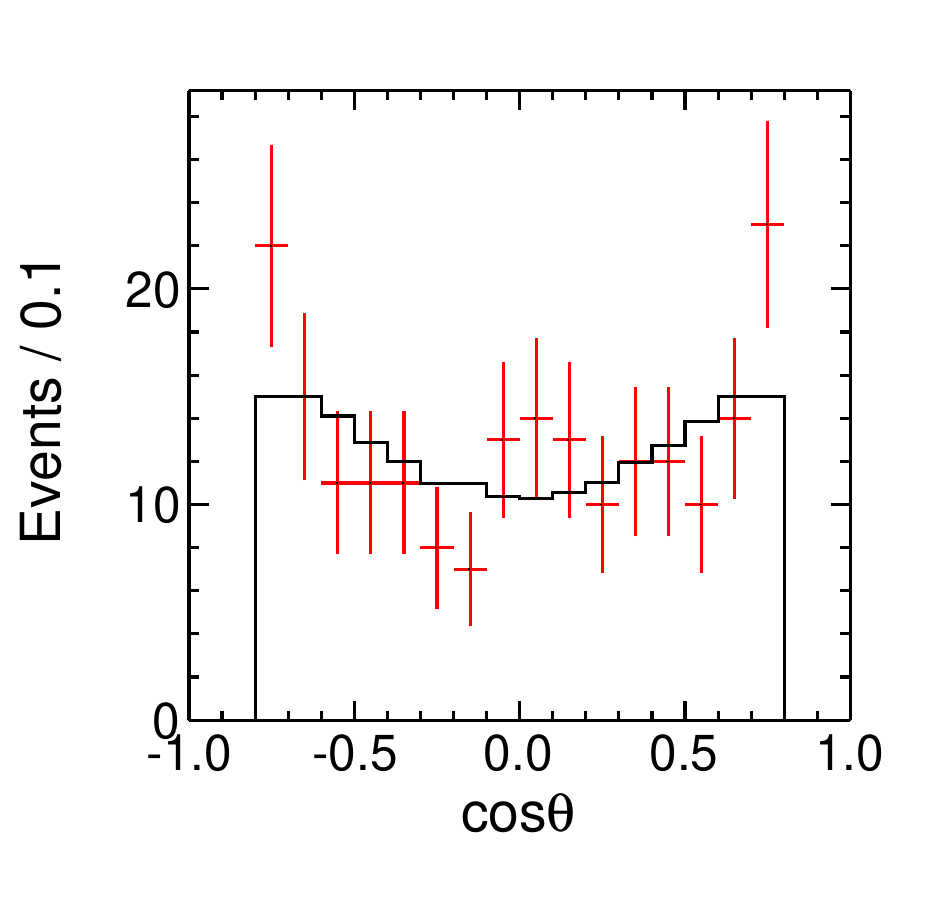}\put(-90,110){\scriptsize $3680.1$~MeV}
  \put(-90,100){\scriptsize $\chi^{2}/{\rm ndf}=0.74$} \\
\includegraphics[angle=0,width=5.0cm]{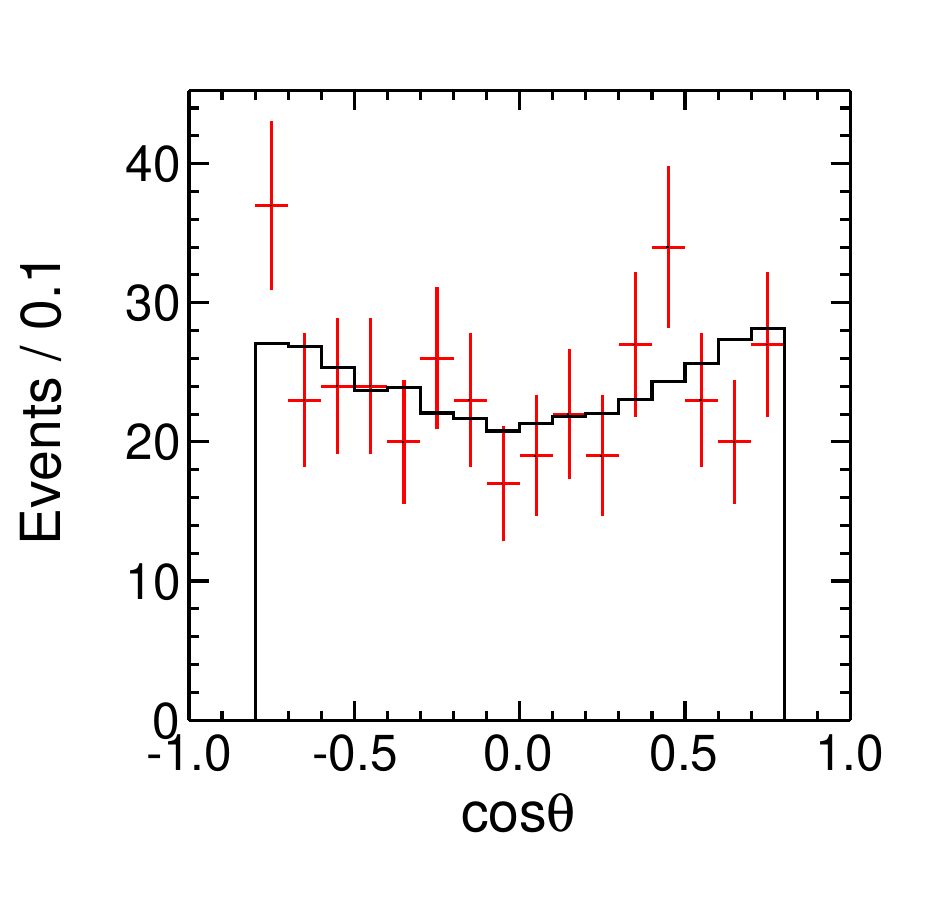}\put(-90,45){\scriptsize $3682.8$~MeV}
  \put(-90,35){\scriptsize $\chi^{2}/{\rm ndf}=0.80$}
\includegraphics[angle=0,width=5.0cm]{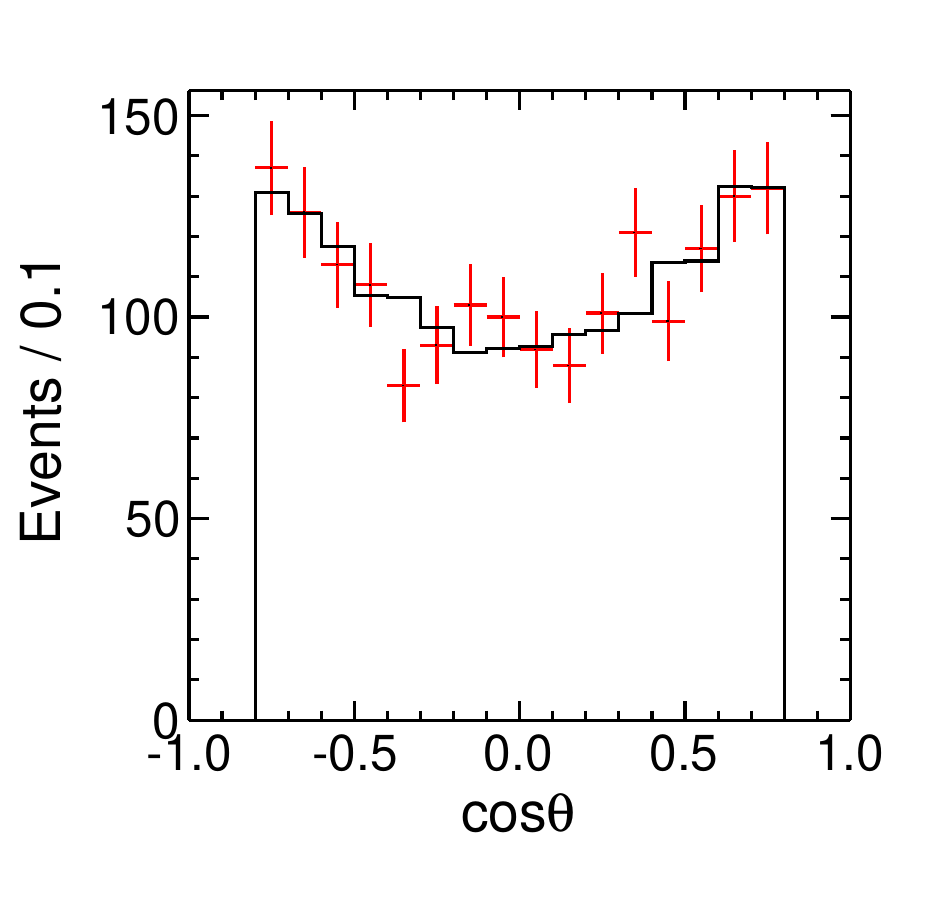}\put(-90,45){\scriptsize $3684.2$~MeV}
  \put(-90,35){\scriptsize $\chi^{2}/{\rm ndf}=0.93$}
\includegraphics[angle=0,width=5.0cm]{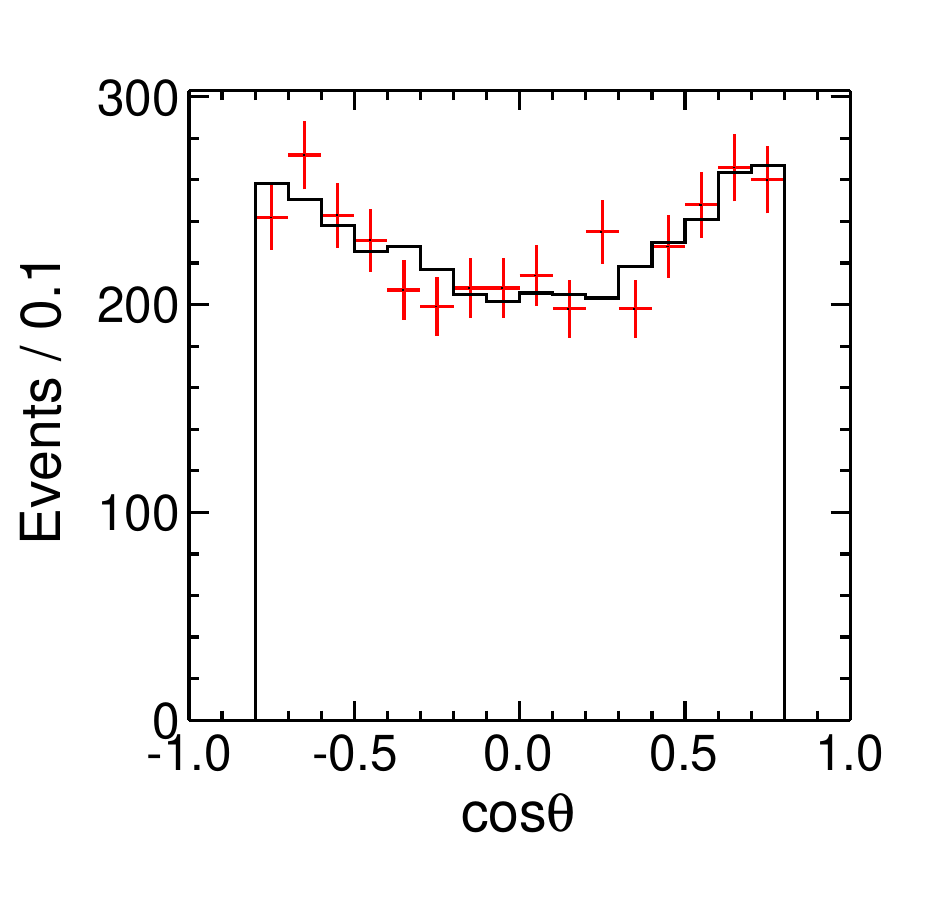}\put(-90,45){\scriptsize $3685.3$~MeV}
   \put(-90,35){\scriptsize $\chi^{2}/{\rm ndf}=0.90$} \\
\includegraphics[angle=0,width=5.0cm]{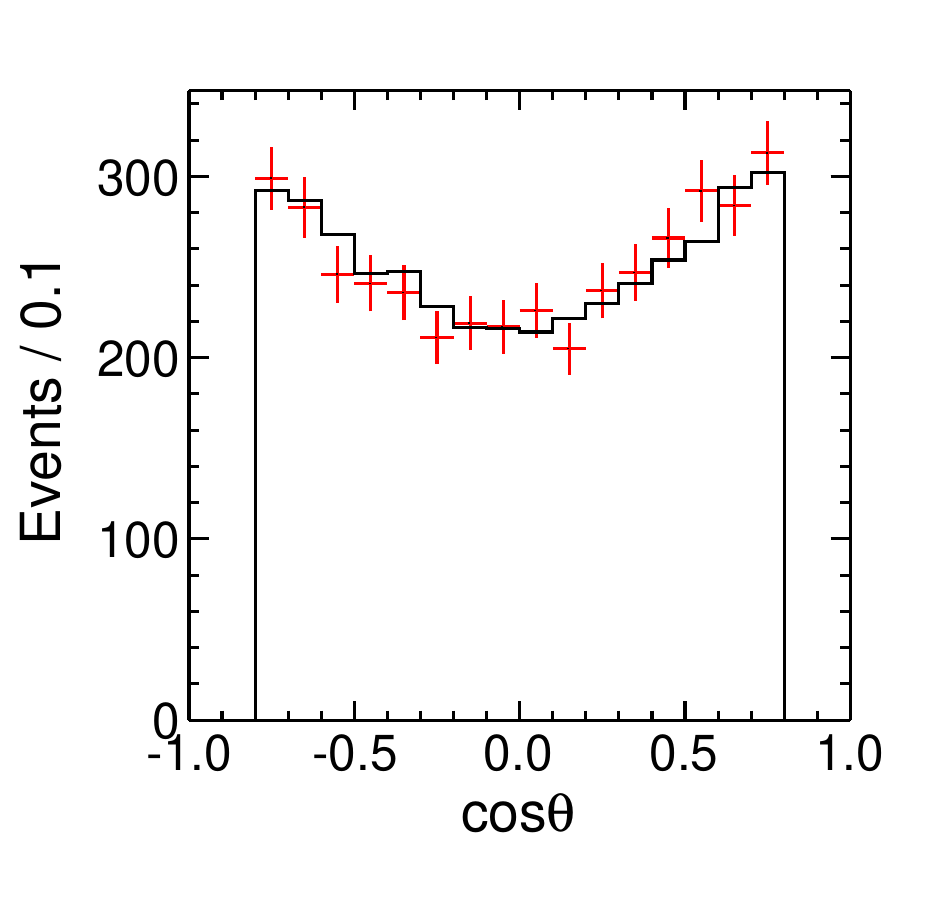}\put(-90,45){\scriptsize $3686.5$~MeV}
   \put(-90,35){\scriptsize $\chi^{2}/{\rm ndf}=0.66$}
\includegraphics[angle=0,width=5.0cm]{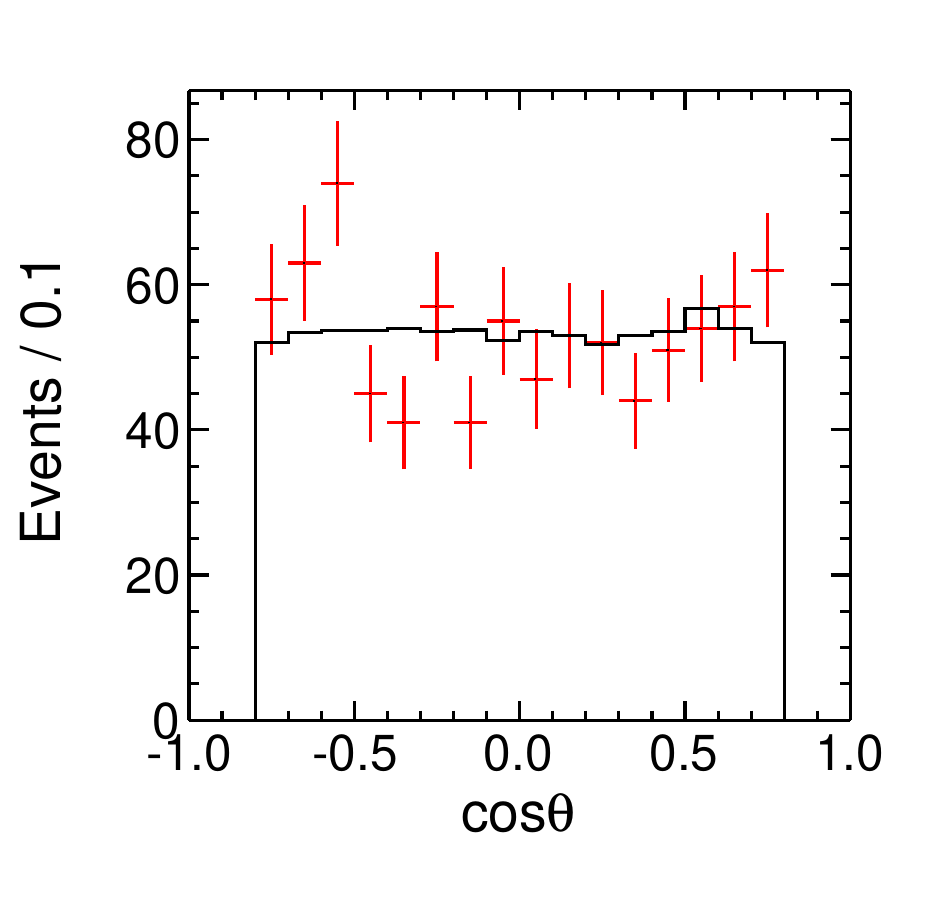}\put(-90,45){\scriptsize $3691.4$~MeV}
   \put(-90,35){\scriptsize $\chi^{2}/{\rm ndf}=1.41$}
\includegraphics[angle=0,width=5.0cm]{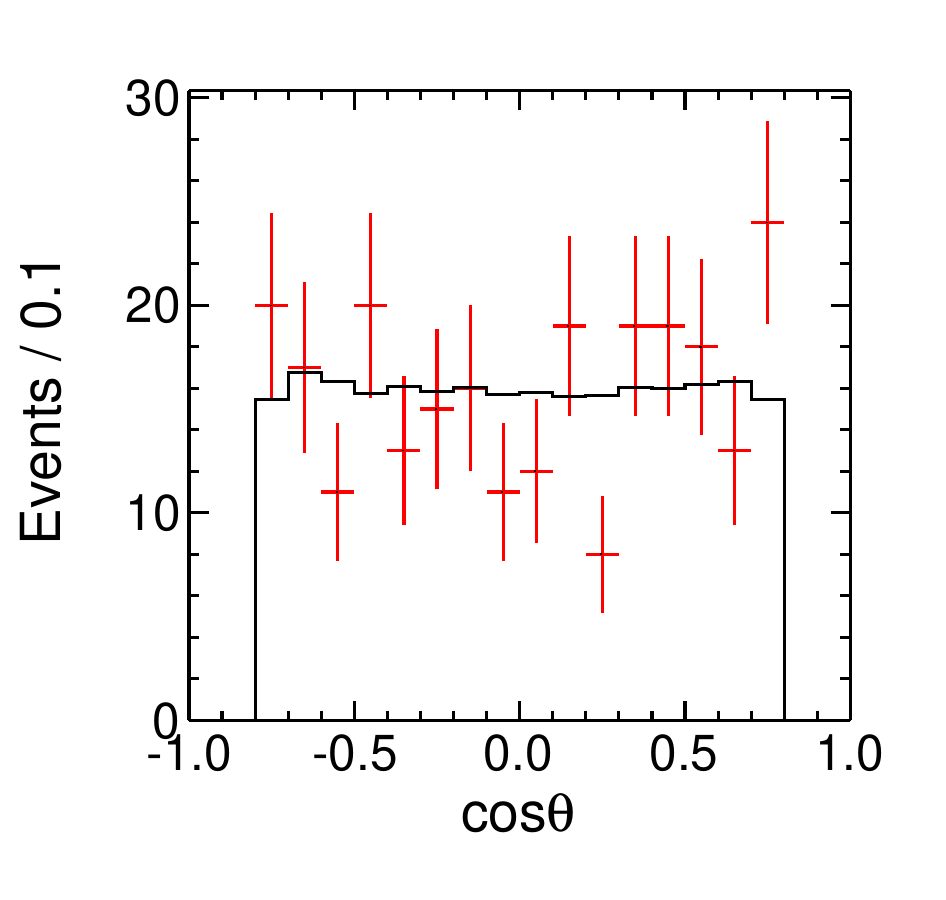}\put(-90,45){\scriptsize $3709.8$~MeV}
    \put(-90,35){\scriptsize $\chi^{2}/{\rm ndf}=1.34$}

\caption{
Comparison of the angular distributions between updated signal MC and data. The red points show the angular distributions of data, and the black solid histograms the angular distributions of signal MC. }
\label{fig_cos}
\end{figure}

The polar angle distributions of the proton in the CM frame are depicted in figure~\ref{fig_cos}, exhibiting a dependence on the CM energies. For a two-body final state, the $\cos\theta$ distribution is well described by the functional form $1+\alpha_{B}\cos^{2}\theta$, where $\alpha_{B}$ is a free parameter. To ensure an accurate simulation of the selection efficiency, the signal MC is systematically tuned to match the experimental data by adjusting the parameter $\alpha_{B}$ across different energy regions. As illustrated in figure~\ref{fig_cos}, the tuned signal MC demonstrates good agreement with experimental data, validating the reliability of the simulation. 

\section{Cross section calculation}

The number of signal events ($N^{\rm obs}$) is extracted by counting events with both tracks having a symmetric momentum window, $[p_{\rm mean}-6\sigma, p_{\rm mean}+6\sigma]$, around the nominal momentum value, $p_{\rm mean}$. The ranges between the blue arrows in figure~\ref{fig_p} represent the momentum window for all energy points. The value of $p_{\rm mean}$ is determined as $\sqrt{s^2/4-m^2_{p}}$. The momentum resolution $\sigma$ is taken from a fit on the momentum distribution from MC simulation with a double-Gaussian function, and $\sigma$ is calculated with $\sigma=\sqrt{f\sigma^{2}_{1}+(1-f)\sigma^2_{2}}$, which is about 15 MeV. The parameter $f$ represents the ratio between the two Gaussian functions, while $\sigma_{1}$ and $\sigma_{2}$ denote the standard deviations of the respective Gaussian functions.
The observed cross section is calculated with 
\begin{equation}
\label{eq_sig}
\sigma^{\rm obs}(s)=\frac{N^{\rm obs}}{\epsilon\mathcal{L}},
\end{equation}
where $\mathcal{L}$ is the integrated luminosity of each dataset.

The detection efficiency $\epsilon$ is estimated with MC samples simulated with {\sc conexc}, and strongly correlated with the input parameterization of the Born cross section \(\sigma^{0}(s)\) as well as the value of \(\alpha_{B}\). To ensure self-consistency, \(\sigma^{0}(s)\) is computed using the parameters extracted from the fit to the observed cross-section line shape. This necessitates an iterative analysis procedure. After several iterations, the results converge, with the relative difference in \(\sigma^{0}(s)\) between the final two iterations found to be below 0.5\%, confirming the stability of the extraction.  
The observed cross sections, integrated luminosities, numbers of observed signal events, and selection efficiencies are summarized in table~\ref{table_xs}.

\begin{table*}[htp]
\small
\caption{The CM energy $\sqrt{s}$~\cite{Abakumova:2011rp}, luminosity $\cal{L}$~\cite{BESIII:2017lkp}, observed event numbers $N^{\rm obs}$, detection efficiencies $\epsilon$, observed cross sections $\sigma^{\rm obs}$ and effective form factors $|G_{\rm eff}|$ at each energy point are summarized. The uncertainties in $N^{\rm obs}$ and $\cal{L}$ are statistical. For $\sigma^{\rm obs}$, the first uncertainties are statistical and the second systematic.}
\label{table_xs}
\begin{center}
\begin{tabular}{|c|c|c|c|c|c|c|c|c|c|c|} \hline
$\sqrt{s}$~(MeV)   &$\cal{L}$~(pb$^{-1}$)    &$N^{\rm obs}$     &$\epsilon$~(\%)   &$\sigma^{\rm obs}$~(pb)  &$|G_{\rm eff}|~(\times10^{-2})$\\ \hline
$3581.5\pm0.1$  &$84.604\pm0.082$  &$76$       &$61.6\pm0.2$    &$1.46\pm0.17\pm0.12$  &$1.45\pm0.09$\\
$3670.2\pm0.1$  &$83.582\pm0.084$  &$83$       &$66.5\pm0.2$    &$1.49\pm0.16\pm0.12$ &$1.31\pm0.08$\\
$3680.1\pm0.1$  &$83.060\pm0.083$  &$141$     &$59.2\pm0.2$    &$2.87\pm0.24\pm0.22$ &$1.30\pm0.08$\\
$3682.8\pm0.1$  &$28.175\pm0.049$  &$350$     &$65.2\pm0.2$    &$19.03\pm1.02\pm1.27$ &$1.29\pm0.08$\\
$3684.2\pm0.1$  &$27.840\pm0.048$  &$1550$    &$64.4\pm0.2$    &$86.49\pm2.20\pm5.28$ &$1.29\pm0.08$\\
$3685.3\pm0.1$  &$25.342\pm0.046$  &$3268$    &$66.3\pm0.2$    &$194.64\pm3.41\pm11.66$ &$1.29\pm0.08$\\
$3686.5\pm0.1$  &$24.481\pm0.045$  &$3581$    &$65.1\pm0.2$    &$224.79\pm3.76\pm13.41$ &$1.29\pm0.08$\\
$3691.4\pm0.1$  &$68.647\pm0.076$  &$744$     &$69.3\pm0.2$    &$15.63\pm0.57\pm0.92$ &$1.28\pm0.08$\\
$3709.8\pm0.1$  &$69.326\pm0.077$  &$191$     &$66.6\pm0.2$    &$4.14\pm0.30\pm0.25$  &$1.26\pm0.08$\\ \hline
\end{tabular}
\end{center}
\end{table*}

\section{Systematic uncertainties}

The systematic uncertainties affecting the observed cross section mainly originate from two categories. The first includes uncertainties related to event selection, MC modeling, the iterative procedure, and the luminosity measurement. These are common to all energy points and are therefore treated as correlated in the cross-section line-shape analysis. The second category comprises uncertainties associated with the momentum window and the polar-angle distribution, which are treated as uncorrelated.

\begin{itemize}
\item 
The systematic uncertainty due to tracking ($\delta_{\rm trk}$) for proton and antiproton is studied using a control sample of $\jpsi\to\prp\prm\pip\pim$ events, and it is found to be 1.0\% per track~\cite{BESIII:2012koo}. Since in this study the final reconstructed particles are $\prp\prm$, a value of 2\% is assigned for the global tracking uncertainty.

\item 
The PID efficiencies of proton and antiproton are studied with the $\psip\to\prp\prm\piz$ control sample, comparing event yields with and without PID requirements. The signal MC sample is generated based on the partial wave analysis result from ref.~\cite{BESIII:2024vqu}. Following the event selection criteria in ref.~\cite{BESIII:2024vqu}, the $\piz$ yield is determined from a fit on the $\gamma\gamma$ invariant mass distribution. A momentum-$\cos\theta$ efficiency matrix for $\prp$ ($\prm$) is constructed, and the  systematic uncertainty at each energy point is estimated using event-fraction weighted bins.
The resulting PID uncertainty ($\delta_{\rm PID}$), treated as fully correlated across energies, is 3.3\% for all energy points.

\item 
The systematic uncertainty ($\delta_{E/P}$) associated with the $E/P$ requirement is also estimated with the $\psip\to\prp\prm\piz$ control sample, following the analysis criteria described in ref.~\cite{BESIII:2024vqu}.
The selection efficiency is determined by comparing the event yields with and without the $E/P$ requirement. A conservative systematic uncertainty of $2.0\%$ is assigned based on the observed data-MC efficiency difference. 

\item 
The $\Delta T$-related systematic uncertainty ($\delta_{\rm T}$) is evaluated by comparing $\Delta T$ distributions between data and MC simulation. To increase the statistical significance, the spectra from all energy points are combined. A double-Gaussian fit to the combined spectrum is performed to extract the efficiencies
in data and MC for the requirement $|\Delta T|<3$ ns. The observed discrepancy is found to be negligible ($\leq 0.1\%$), resulting in a correspondingly negligible systematic uncertainty.

\item  
The systematic uncertainty ($\delta_{\theta_{\prp\prm}}$) induced by the requirement of $\theta_{\prp\prm}$ is evaluated by comparing the cross section values obtained varying the cut conditions~(with and without 178${^\circ}$ \textless $\theta_{\prp\prm}$ \textless 180 ${^\circ}$) in $\psip$ data. The resulting uncertainty is determined to be 0.3\%.

\item 
The systematic uncertainty ($\delta_{\rm model}$) arising from the MC model is estimated by considering the efficiency difference (0.2\%) observed between the last two iterations.
Additionally, the intrinsic uncertainty of the initial state radiation (ISR) function implemented in {\sc conexc} is 0.5\%~\cite{Ping:2013jka}, which is calculated by adding all systematic uncertainty in quadrature.
Thus, the systematic uncertainty attributed to the MC model is determined as 0.5\%.

\item 
The momentum window is varied with $\pm 1\sigma$ to estimate the systematic uncertainties ($\delta_{P}$) due to the momentum window cut. The largest difference in the cross section is taken as the systematic uncertainty for each energy point.

\item 
The systematic uncertainty ($\delta_{\alpha_{B}}$) arising from the imperfect simulation of the angular distribution is estimated by varying the parameter $\alpha_{B}$ by one standard deviation. The largest observed difference in the cross section is assigned as the systematic uncertainty due to the angular distribution for each energy point.

\item 
The iterative analysis employs parameters obtained from fitting the cross section line shape to configure subsequent MC generations. To avoid bias, the fit procedure must exactly match the {\sc conexc} generator's calculation. The associated systematic uncertainty ($\delta_{\rm iter}$) is quantified as the maximum cross section difference (3\%) between the generator and the fit results~\cite{Wang:2023kxu}.

\item 
The integrated luminosities of the data sets are measured using di-photon events, with an uncertainty ($\delta_{\cal{L}}$) of about 1.0\%~\cite{BESIII:2017lkp} which is propagated to the cross section measurements. 

\end{itemize}
Table~\ref{table_sys_fit} lists the individual uncertainties and the total uncertainty ($\delta_{\rm tot}$), which is the sum of  all of the individual systematic uncertainties in quadrature.

\begin{table*}[htp]
\small
\caption{Summary of relative systematic uncertainties (in unit of \%) in the cross section measurement.}
\label{table_sys_fit}
\begin{center}
\begin{tabular}{|c|c|c|c|c|c|c|c|c|c|c|} \hline
$E_{\rm CM}$~(MeV)   &$\delta_{\rm trk}$ &$\delta_{\rm PID}$ &$\delta_{E/p}$&$\delta_{\rm \theta_{\prp\prm}}$ &$\delta_{\rm model}$ &$\delta_{P}$ &$\delta_{\alpha_{B}}$  &$\delta_{\rm iter}$ &$\delta_{\mathcal{L}}$ &$\delta_{\rm tot}$  \\ \hline
$3581.5$ &$2$             &$3.3$   &$2$   &$0.3$   &$0.5$ &$1.2$ &$5.9$  &$3$ &$1$ &$8.4$ \\
$3670.2$ &$2$             &$3.3$   &$2$   &$0.3$   &$0.5$ &$0.1$ &$5.9$  &$3$ &$1$ &$8.3$ \\
$3680.1$ &$2$             &$3.3$   &$2$   &$0.3$   &$0.5$ &$0.7$ &$4.8$ &$3$ &$1$ &$7.6$ \\
$3682.8$ &$2$             &$3.3$   &$2$   &$0.3$   &$0.5$ &$0.5$ &$3.1$  &$3$ &$1$ &$6.7$ \\
$3684.2$ &$2$             &$3.3$   &$2$   &$0.3$   &$0.5$ &$0.4$ &$1.7$  &$3$ &$1$ &$6.1$ \\
$3685.3$ &$2$             &$3.3$   &$2$   &$0.3$   &$0.5$ &$0.2$ &$1.2$  &$3$ &$1$ &$6.0$ \\
$3686.5$ &$2$             &$3.3$   &$2$   &$0.3$   &$0.5$ &$0.2$ &$1.1$  &$3$ &$1$ &$6.0$ \\
$3691.4$ &$2$             &$3.3$   &$2$   &$0.3$   &$0.5$ &$0.3$ &$0.1$  &$3$ &$1$ &$5.9$ \\
$3709.8$ &$2$             &$3.3$   &$2$   &$0.3$   &$0.5$ &$0.1$ &$1.1$  &$3$ &$1$ &$6.0$ \\ \hline
\end{tabular}
\end{center}
\end{table*}

\section{Fit results for the line shape}\label{sec_fit}

The parametrization of $\sigma^{0}(s)$ is described in equation~\ref{eq:xs1}. 
In the fit, $\Gamma_{ee}$ is the partial width of the $\psip\to\elp\elm$ decay including the vacuum polarization, from the PDG~\cite{ParticleDataGroup:2024cfk}.
Convolved with the ISR function and a Gaussian function to account for initial-state radiation effects ISR~$(F(x,s))$ and beam energy spread effects~$(G(\sqrt{s}-\sqrt{s^\prime}))$, the theoretical observed cross section is calculated as
\begin{eqnarray}
\label{eq_BWG}
\sigma^{\prime\prime}(s) = \int^{\sqrt{s}+nS_{E}}_{\sqrt{s}-nS_{E}} G(\sqrt{s}-\sqrt{s^\prime}) d\sqrt{s^{\prime}} \int \limits_{0}^{x_f}  F(x,s) \sigma^{0}(s(1-x)) dx,
\end{eqnarray}
where $x_f$ is the upper limit of $x=1-\frac{s^\prime}{s}$, corresponding to the maximum energy of the radiation,  $S_{E}$ is the CM energy spread determined by the accelerator operating conditions,
$\sqrt{s^\prime}$ is the measured
 invariant mass of the final state  after losing energy due to radiation, and  $s(1-x)$ is the CM energy squared after considering ISR.
The minimization function is built with the  factorized minimization method separating the correlated and uncorrelated systematic uncertainties, and the effective variance-weighted least squares method~\cite{Roe_chisq} is used by
\begin{equation} \label{eq:chisq}
\chi^{2} = \sum^{9}_{i=1} \frac{ \left[\sigma^{\rm obs}_{i}-f\sigma^{\prime\prime}(W_{i})\right]^{2} }{ (\Delta\sigma^{\rm obs}_{i})^{2} + \left[\Delta W_{i}\cdot \frac{d\sigma^{\prime\prime}(W)}{dW}\right]^{2} } + \left(\frac{1-f}{\Delta f}\right)^2, 
\end{equation}
where $W=\sqrt{s}$, $\Delta W$ is the uncertainty of $W$ measured by the BEMS, $f$ is the normalization factor, and $\Delta f$ is defined as the total correlated uncertainty and obtained by combining all correlated systematic uncertainties in quadrature.
The term $\Delta\sigma^{\rm obs}$ is the combined statistical and uncorrelated systematic uncertainties of $\sigma^{\rm obs}$.
To improve the regression speed, instead of making a two-fold integration, the analytical formula proposed in ref.~\cite{Wang:2023kxu} is used. 

\begin{figure}[htbp]\center
\includegraphics[angle=0,width=10cm, height=9cm]
{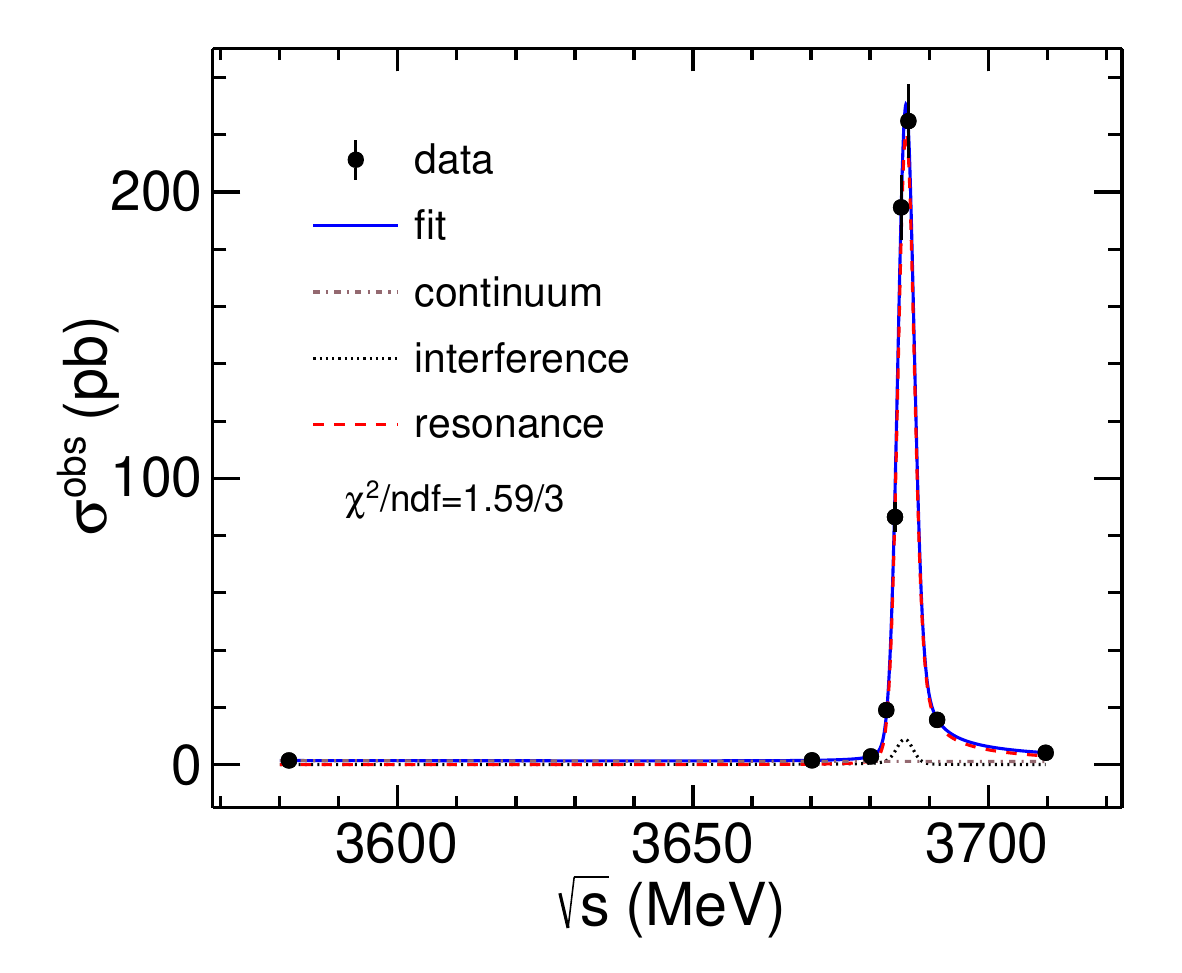}\put(-100,70){(a)} \\
\includegraphics[angle=0,width=10cm, height=9cm]{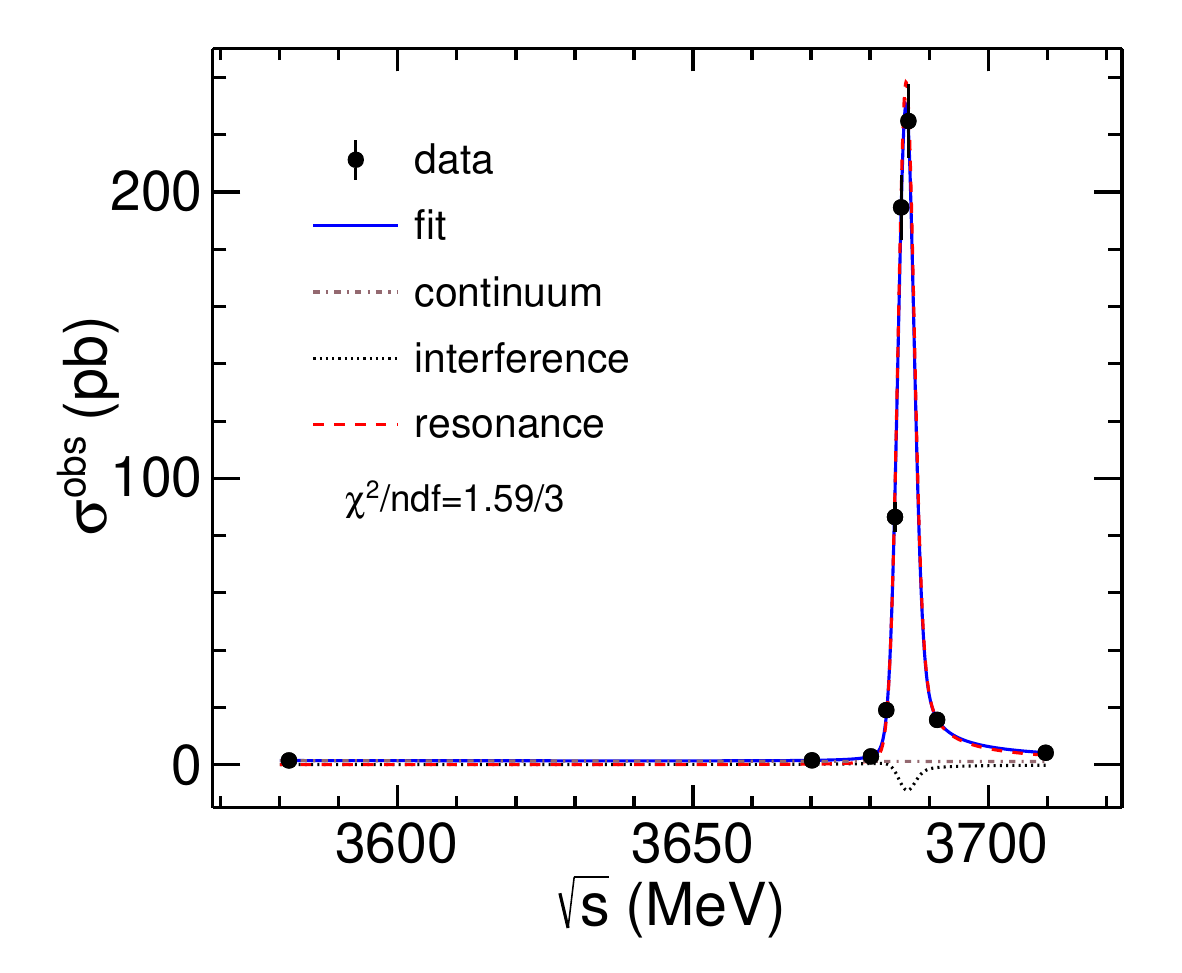}\put(-100,70){(b)}
\caption{Fit results on the observed cross section of $\elp\elm\to\prp\prm$, where (a) is the positive solution and (b) is the negative solution. The black points with error bars show the observed cross sections with statistical uncertainties and uncommon systematic uncertainties. The blue solid curves denote the total theoretical cross section, while the red dashed, brown dot-dashed and black dotted lines denote the contributions from resonance, continuum and interference, respectively. The (a) plot is the constructive result, while (b) is the destructive result.}
\label{fig_fit}
\end{figure}

The fitted parameters are presented in table~\ref{tab_result} which uncertainties include both statistical and systematic contributions. Since the experimental observables depend on the moduli of the amplitudes, two distinct solutions are obtained with identical fit quality ($\chi^2$). The corresponding fit curves are shown in figure~\ref{fig_fit}.

\begin{table}[htp]
\caption{Results of parameters in the cross section line shape fit.}
\label{tab_result}
\centering
\begin{tabular}{|c|c|c|c|c|c|c|c|c|} \hline
Parameter                  &Positive             &Negative             \\\hline
$\Phi_{g,EM} (^\circ)$     &$105.9\pm9.4$        &$-105.3\pm9.1$       \\
$M_{\psi}$~(MeV$/c^{2}$)   &$3685.99\pm0.08$     &$3685.99\pm0.08$        \\
$\mathcal{C}$              &$16.02\pm1.25$       &$16.63\pm1.23$        \\
$G_0$ (GeV$^4$)            &$2.38\pm0.15$        &$2.38\pm0.15$      \\
$\mathcal{B}$ ($10^{-4}$)  &$3.21\pm0.13$        &$3.47\pm0.19$\\\hline
\end{tabular}
\end{table}

The branching fraction is determined with the parameters in table~\ref{tab_result} via
\begin{eqnarray}\label{eq_Br}
\mathcal{B}=\frac{\Gamma_{\prp\prm}}{\Gamma_{ee}}\mathcal{B}_{ee}
=\beta C(\frac{G_{0}}{M_{\psi}^4})^2(1+\frac{2m^2_{\prp}}{M_{\psi}^2})|1+\mathcal{C}e^{i\Phi_{g,\gamma}}|^2\mathcal{B}_{ee},
\end{eqnarray}
where $\mathcal{B}_{ee}$ is the branching fraction of the $\psip\to\elp\elm$ decay. To take into account the correlations among all parameters in the cross section line shape fit, the branching fraction $\mathcal{B}$ and its associated uncertainty are determined by sampling from multivariate Gaussian distributions, using the correlation coefficients from the nominal fit. The central value of $\mathcal{B}$ is derived from its distribution, as listed in table~\ref{tab_result}. 

The systematic uncertainties from the fixed parameters ($\Gamma$, $\Gamma_{ee}$) are evaluated by varying them by $\pm1\sigma$. The largest difference listed in table~\ref{tab_uncertainty} is considered and is added quadratically in the final result shown in table~\ref{tab_result}.

\begin{table}[htp]
\caption{Systematic uncertainties (in \%) for every parameter obtained by varying $\Gamma_{\psi}$ and $\Gamma_{ee}$.}
\label{tab_uncertainty}
\centering
\begin{tabular}{|c|c|c|c|c|c|c|c|c|c|c|c|c|} \hline
 Parameter  &$M_{\psi}$~(MeV$/c^{2}$)  &$\Phi_{g,\gamma} (^\circ)$  &$S_{E}$  &$\mathcal{C}$ &$G_{0}$ &$\mathcal{B}$ \\ \hline
$\delta~(\Phi_{g,\gamma}< 0)$ &$0.0$  &$0.4$  &$1.1$  &$3.2$ &$0.1$ &$2.4$ \\ \hline
$\delta~(\Phi_{g,\gamma}> 0)$ &$0.0$  &$0.4$  &$1.0$  &$3.3$ &$0.1$ &$2.3$ \\ \hline
\end{tabular}
\end{table}

\section{Effective  form factors}

\begin{figure}[htbp]
\begin{center}
\includegraphics[angle=0,width=10cm, height=9cm]{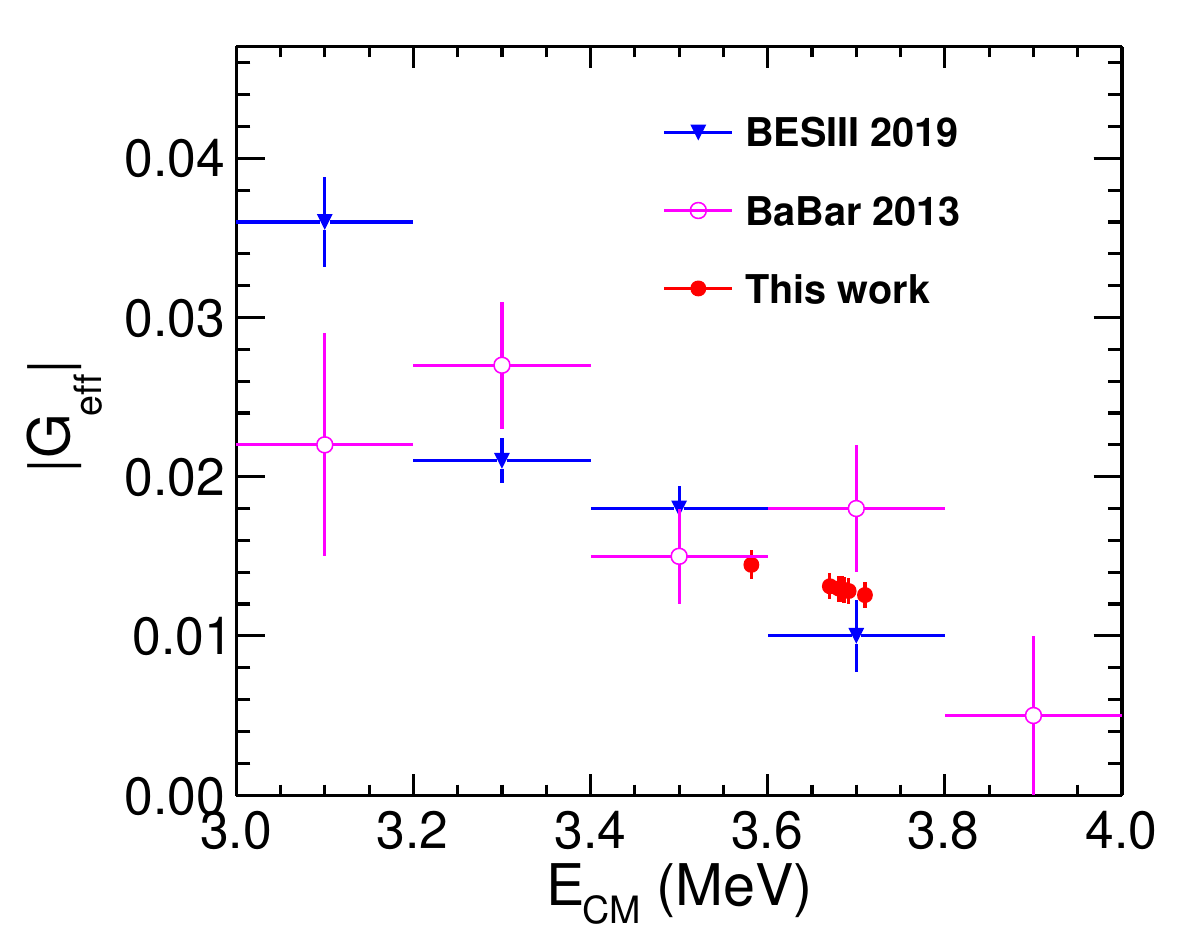}
\end{center}
\caption{The effective proton form factors calculated in this work (red dots with error bars) compared with the previous results from BESIII (blue triangles) and BaBar (magenta circles).
}
\label{fig_FF}
\end{figure}

From refs.~\cite{BESIII:2019hdp,BESIII:2021rqk,Huang:2021xte}, the Born cross section of $\elp\elm\to\prp\prm$ in the continuum region is calculated through 
\begin{eqnarray}\label{eq_xs_cont}
    \sigma^{0}_{\rm cont.}=\frac{4\pi\alpha^2\beta C}{3s}(1+\frac{2m^2_{p}}{s})|G_{\rm eff}|^2\,,
\end{eqnarray}
where $\beta=\sqrt{1-4m^2_{p}/s}$ is the phase space factor, $C$ is the Coulomb effect between proton and anti-proton, $|G_{\rm eff}|$ is the effective form factor. 
With the result of $G_{0}$ extracted from the fit,
the effective EMFFs ($|G_{\rm eff}|$) around $\psip$ are calculated and the obtained results are shown in table~\ref{table_xs}. The comparison with the previous results is shown in figure~\ref{fig_FF}. Our result shows improved precision and is consistent within 3$\sigma$ with the previous BESIII~\cite{BESIII:2019hdp,BESIII:2021rqk} and BaBar~\cite{BaBar:2015lgl} results. 


\section{Summary}

In this paper, we present the measurement of the cross-section line shape for the process $\elp\elm\to\prp\prm$ around the $\psip$ resonance.
The relative phase between the strong and EM mechanisms is measured as two solutions, 
$\Phi_{g,\gamma}=(105.9\pm9.4)^\circ$ and $\Phi_{g,\gamma}=(-105.3\pm9.1)^\circ$.
Correspondingly, the branching fraction of $\psi(2S)\to \prp\prm$ are determined as 
$\mathcal{B}=(3.21\pm0.13)\times10^{-4}$ and $\mathcal{B}=(3.47\pm0.19)\times10^{-4}$.
The uncertainties include both statistic and systematic contributions.
Both phase solutions indicate an orthogonal relation between the strong and the electromagnetic amplitudes around the $\psip$ state. However, the fundamental mechanism behind this phenomenon remains to be elucidated theoretically.
For the branching fraction, both our results are larger but still consistent within three standard deviations compared to the PDG value $\mathcal{B}=(2.94\pm0.08)\times10^{-4}$~\cite{ParticleDataGroup:2024cfk}. This discrepancy may arise from various reasons, such as the unaccounted interference between the resonance and the continuum amplitudes in the previous results. Additionally, such interference can shift the peak position from its true value, leading to deviations in the branching fraction measurement. In this case, the scanning method demonstrates superiority for the precise measurement.
The effective proton form factors in the $\psip$ energy interval are  determined based on the parameters extracted from the cross-section line shape, and the results are consistent with the previous BESIII and BaBar results.

\section*{Acknowledgments}

The BESIII Collaboration thanks the staff of BEPCII (https://cstr.cn/31109.02.BEPC) and the IHEP computing center for their strong support. This work is supported in part by National Key R\&D Program of China under Contracts Nos. 2023YFA1606000, 2023YFA1606704; National Natural Science Foundation of China (NSFC) under Contracts Nos. 12105100, 11635010, 11935015, 11935016, 11935018, 12025502, 12035009, 12035013, 12061131003, 12192260, 12192261, 12192262, 12192263, 12192264, 12192265, 12221005, 12225509, 12235017, 12342502, 12361141819;  Beijing Natural Science Foundation (BJNSF) under Contract No. JQ22002; the Chinese Academy of Sciences (CAS) Large-Scale Scientific Facility Program; the Strategic Priority Research Program of Chinese Academy of Sciences under Contract No. XDA0480600; CAS under Contract No. YSBR-101; 100 Talents Program of CAS; The Institute of Nuclear and Particle Physics (INPAC) and Shanghai Key Laboratory for Particle Physics and Cosmology; ERC under Contract No. 758462; German Research Foundation DFG under Contract No. FOR5327; Istituto Nazionale di Fisica Nucleare, Italy; Knut and Alice Wallenberg Foundation under Contracts Nos. 2021.0174, 2021.0299, 2023.0315; Ministry of Development of Turkey under Contract No. DPT2006K-120470; National Research Foundation of Korea under Contract No. NRF-2022R1A2C1092335; National Science and Technology fund of Mongolia; Polish National Science Centre under Contract No. 2024/53/B/ST2/00975; STFC (United Kingdom); Swedish Research Council under Contract No. 2019.04595; U. S. Department of Energy under Contract No. DE-FG02-05ER41374.

\bibliography{reference_v2}

\clearpage
\appendix
M.~Ablikim$^{1}$\BESIIIorcid{0000-0002-3935-619X},
M.~N.~Achasov$^{4,d}$\BESIIIorcid{0000-0002-9400-8622},
P.~Adlarson$^{81}$\BESIIIorcid{0000-0001-6280-3851},
X.~C.~Ai$^{87}$\BESIIIorcid{0000-0003-3856-2415},
C.~S.~Akondi$^{31A,31B}$\BESIIIorcid{0000-0001-6303-5217},
R.~Aliberti$^{39}$\BESIIIorcid{0000-0003-3500-4012},
A.~Amoroso$^{80A,80C}$\BESIIIorcid{0000-0002-3095-8610},
Q.~An$^{77,64,\dagger}$,
Y.~H.~An$^{87}$\BESIIIorcid{0009-0008-3419-0849},
Y.~Bai$^{62}$\BESIIIorcid{0000-0001-6593-5665}
% ...
\\
\vspace{0.2cm}
(BESIII Collaboration)\\
\vspace{0.2cm} {\it
$^{1}$ Institute of High Energy Physics, Beijing 100049, People's Republic of China\\
% ...
$^{4}$ Budker Institute of Nuclear Physics SB RAS (BINP), Novosibirsk 630090, Russia\\
% ...
$^{31}$ INFN Sezione di Ferrara, (A)INFN Sezione di Ferrara, I-44122, Ferrara, Italy; (B)University of Ferrara, I-44122, Ferrara, Italy\\
% ...
$^{39}$ Johannes Gutenberg University of Mainz, Johann-Joachim-Becher-Weg 45, D-55099 Mainz, Germany\\
% ...
$^{62}$ Southeast University, Nanjing 211100, People's Republic of China\\
% ...
$^{64}$ State Key Laboratory of Particle Detection and Electronics, Beijing 100049, Hefei 230026, People's Republic of China\\
% ...
$^{77}$ University of Science and Technology of China, Hefei 230026, People's Republic of China\\
% ...
$^{80}$ University of Turin and INFN, (A)University of Turin, I-10125, Turin, Italy; (B)University of Eastern Piedmont, I-15121, Alessandria, Italy; (C)INFN, I-10125, Turin, Italy\\
$^{81}$ Uppsala University, Box 516, SE-75120 Uppsala, Sweden\\
% ...
$^{87}$ Zhengzhou University, Zhengzhou 450001, People's Republic of China\\

\vspace{0.2cm}
$^{\dagger}$ Deceased\\
% ...
$^{d}$ Also at the Novosibirsk State University, Novosibirsk, 630090, Russia\\
% ...

}

\end{document}